\documentclass[twocolumn,amsmath,amssymb,pra,longbibliography,floatfix]{revtex4-2}
\usepackage{graphicx}
\usepackage[table]{xcolor}
\usepackage{amsmath,amsfonts,amssymb,amsthm}
\usepackage{fancyhdr}
\usepackage{mathrsfs}
\usepackage{epstopdf}
\usepackage{setspace}
\usepackage{bm}
\usepackage[colorlinks=true, citecolor=blue, linkcolor=blue]{hyperref}
\usepackage{xcolor}
\usepackage{dcolumn}
\usepackage{tikz}
\usepackage{tikz}
\usetikzlibrary{arrows.meta, positioning, shapes.multipart}
\usepackage{graphicx}
\usepackage{array}
\usepackage{multirow}
\usetikzlibrary{quantikz}
\graphicspath{{figures/}}
\usepackage{pifont} 
\usepackage{siunitx} 
\newcommand{\xmark}{\ding{55}} 
\definecolor{hqsorange}{RGB}{217,120,21}
\definecolor{hqsgreen}{RGB}{67,174,85}
\definecolor{hqsblue}{RGB}{64,127,193}
\definecolor{hqstext}{RGB}{164,94,42}
\definecolor{hqsred}{RGB}{235,68,66}
\definecolor{inactivecolor}{gray}{0.92}
\newcommand{\inactivestar}{\textcolor{inactivecolor}{\ding{72}}}
\newcommand{\starzero}{\inactivestar\inactivestar\inactivestar\inactivestar\inactivestar} 
\newcommand{\starone}{\ding{72}\inactivestar\inactivestar\inactivestar\inactivestar} 
\newcommand{\startwo}{\ding{72}\ding{72}\inactivestar\inactivestar\inactivestar} 
\newcommand{\starthree}{\ding{72}\ding{72}\ding{72}\inactivestar\inactivestar} 
\newcommand{\starfour}{\ding{72}\ding{72}\ding{72}\ding{72}\inactivestar} 
\newcommand{\starfive}{\ding{72}\ding{72}\ding{72}\ding{72}\ding{72}} 

\begin{document}
\title{What is a good use case for quantum computers?}


\author{Michael Marthaler, Peter Pinski, Vladimir Rybkin, Iris Schwenk, Pascal Stadler, Marina Walt} 
\affiliation{HQS Quantum Simulations GmbH, Rintheimer Str. 23, 76131 Karlsruhe, Germany}

\begin{abstract}
Identify, Transform, Benchmark, Show Quantum Advantage (ITBQ): Evaluating use cases for quantum computers.
We introduce a four-step framework for assessing quantum computing applications -- from identifying relevant industry problems to demonstrating quantum advantage -- addressing steps often overlooked in the literature, such as rigorous benchmarking against classical solutions and the challenge of translating real-world tasks onto quantum hardware.
Applying this framework to cases like NMR, multireference chemistry, and radicals reveals both significant opportunities and key barriers on the path to practical advantage.
Our results highlight the need for transparent, structured criteria to focus research, guide investment, and accelerate meaningful quantum progress.
\end{abstract}

\maketitle

\section{Introduction}

Quantum computing hardware has progressed substantially in the last few years. Several quantum computing hardware providers can relatively consistently achieve two qubit gates with more than 99\% Fidelity. Under laboratory conditions, even 99.9\% has been achieved in quite a few architectures, and at least one cloud-accessible device with these remarkable fidelities exists (Quantinuum). 
This all makes it even more important to drive use cases forward. A vast amount of work has been dedicated to testing various algorithms and approaches in the last few years~\cite{Preskill2018,Blekos2024,Biamonte2017,Garcia2022,Bauer2020,McArdle2020} and it is now all the more important to assess the state and quality of the existing ideas for use cases. 

In this work, we want to discuss the quality and current state of some exemplary use cases using four interdependent criteria:
\begin{itemize}
 \item Identify industry problem,
 \item Transform to quantum,
 \item Get the job done without the quantum computer,
 \item Show quantum advantage.
\end{itemize}

In the literature on quantum computing, discussions often focus on two main criteria: identifying an industry problem and demonstrating quantum advantage. This approach is based on the belief that there should be a genuinely interesting problem to solve. Ideally, this problem should be relevant to industry, but at the very least, it should have some academic significance. It should not merely be a problem created to showcase quantum advantage. Additionally, evaluations of use cases often include arguments or rigorous algorithmic proofs to explore how and when quantum advantage can be achieved.

However, it is often noticable that combining just these two criteria in a completely satisfactory way remains quite challenging. As we will discuss later, there is a vast number of studies on problems that are without any doubts of great industrial relevance, but lack substantiated arguments on how to achieve quantum advantage. Concurrently, it is highly nontrivial to connect algorithms with strictly provable quantum advantage to actual applications. Most prominently, this is the case for sampling experiments which are certainly of crucial importance as a fundamental performance benchmark, but until recently~\cite{zhang2023protecting} lacked any connection to applications. 

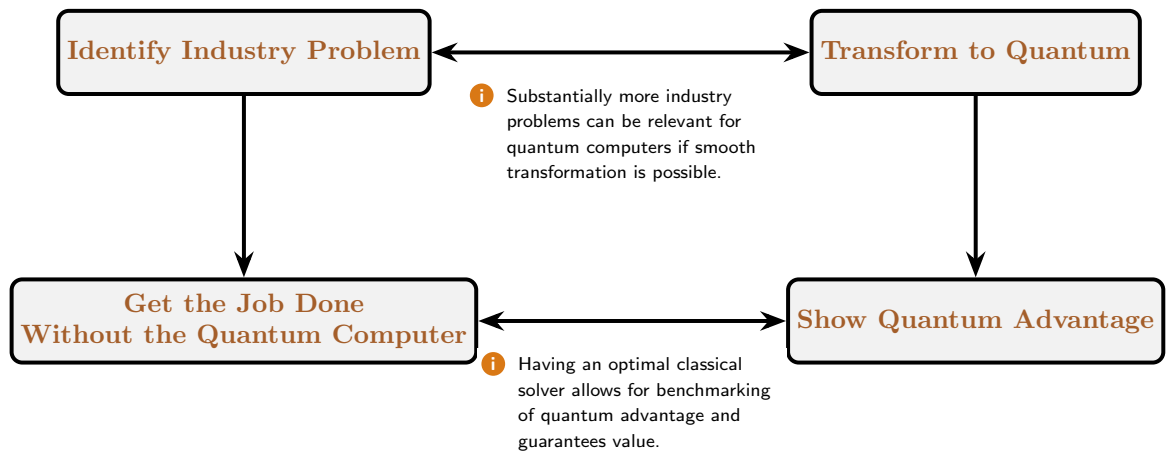
\begin{figure*}[t]
  \centering
  \begin{tikzpicture}[>=Stealth, node distance=5cm, font=\sffamily,line width=1.5pt]
  
  \node (industry) [draw, rounded corners, align=center, minimum width=2.6cm, minimum height=1.1cm, fill=gray!10, text=hqstext, font=\bfseries] {Identify Industry Problem};
  \node (transform) [draw, rounded corners, align=center, right=of industry, minimum width=2.6cm, minimum height=1.1cm, fill=gray!10, text=hqstext, font=\bfseries] {Transform to Quantum};
  
  \draw[<->, thick, line width=1.5pt] (industry) -- (transform)
  node[midway, fill=white, inner sep=1.5pt, below=10pt] {%
    \scriptsize
    \begin{tabular}{@{}l@{}}
      \tikz[baseline=(i.base)]
        \node[circle, fill=hqsorange, inner sep=0pt, minimum size=1.2em] (i)
            {\sffamily\bfseries\textcolor{white}{i}};~
      \parbox[t]{3.5cm}{\raggedright
        \setstretch{1.2} 
        Substantially more industry problems can be relevant for quantum computers if smooth transformation is possible.
      }
    \end{tabular}%
};

  \node (classical) [draw, rounded corners, align=center, below=2.4cm of industry, minimum width=2.6cm, minimum height=1.1cm, fill=gray!10,text=hqstext, font=\bfseries] {Get the Job Done\\ Without the Quantum Computer};
  \node (quantum) [draw, rounded corners, align=center, below=2.4cm of transform, minimum width=2.6cm, minimum height=1.1cm, fill=gray!10,text=hqstext, font=\bfseries] {Show Quantum Advantage};
  
  \draw[<->, thick, line width=1.5pt] (classical) -- (quantum)
      node[midway, fill=white, inner sep=1.5pt, below=10pt] {%
        \scriptsize
        \begin{tabular}{@{}l@{}}
          \tikz[baseline=(i.base)]
            \node[circle, fill=hqsorange, inner sep=0pt, minimum size=1.2em] (i)
                {\sffamily\bfseries\textcolor{white}{i}};~
          \parbox[t]{3.5cm}{\raggedright
            \setstretch{1.2} 
            Having an optimal classical solver allows for benchmarking of quantum advantage and guarantees value.
          }
        \end{tabular}%
    };
      
  \draw[->, thick,line width=1.5pt] (industry) -- (classical)
      node[midway, right=10pt, fill=white, align=center, inner sep=2pt, font=\scriptsize]
      {};

    \draw[->, thick, line width=1.5pt] (transform) -- (quantum)
    node[midway, right=10pt, fill=white, align=center, inner sep=2pt, font=\scriptsize]
    {};
  
  \end{tikzpicture}
  \caption{Steps to assess quantum computing use cases: \textbf{I}dentify, \textbf{T}ransform, \textbf{B}enchmark, Show \textbf{Q}uantum Advantage (ITBQ).}
  \label{fig:criteria}
  \end{figure*}  

Apart from reviewing some important historic use cases and three use cases from our own team, we intend to add two important criteria for assessing quantum computing use cases. One criterion -- \textit{Transform to Quantum} -- is about the necessity to consider the step of transforming the actual input as you get it from an end user into the formulation required for the quantum computer. A key aspect here is that only a small subset of the problem might benefit from using the quantum computer and needs to be efficently extracted from a larger problem. When evaluating the quality of current use cases, particular attention should be given to the state of research and the availability of software to perform this transformation step. This step can present a significant barrier to effective use case implementation if the software required to identify and extract the most relevant subproblem is not available. At the same time, it can serve as the crucial step to connect otherwise abstract quantum algorithms to real end users.

The second criterion we intend to add is the need to invest as much effort to solve the given problem with classical computers as into the research on quantum computing. We call this criteria  \textit{Get the Job Done Without the Quantum Computer}. Of course, scientific research often has limited resources, so it might not always be possible to fully optimize classical methods before exploring the quantum ones. Nevertheless, at some stage in the process, it’s important to consider the status and availability of optimized classical solutions when judging any work on a quantum use case. As an example, one of the first claims advanced for an andvantage in using quantum computers was based on a benchmark performed by a D-Wave quantum computer~\cite{mcgeoch2013experimental} versus classical CPLEX solver, a standard tool for combinatorial optimization problems. However, this example did not stand up to scrutiny in the long run, since CPLEX is not the best possible solver for the type of problem addressed by the D-Wave architecture~\cite{ronnow2014defining}. In another recent instance, an impressive large-scale simulation of spin dynamics on a quantum computer was reproduced by several groups using classical computers in a matter of weeks~\cite{mauron2025challengingquantumadvantage}. Therefore, the question if the problem has been compared to optimized classical solvers is always quite important to assess a quantum computing use case.

These four steps -- \textbf{I}dentify, \textbf{T}ransform, \textbf{B}enchmark, Show \textbf{Q}uantum Advantage (ITBQ) -- are obviously also interconnected (Fig.~\ref{fig:criteria}). Once the relevant industry problem has been identified, one needs to explore if optimized classical solvers are also available. When classical solvers are sufficient to solve the identified industry problem, the quantum advantage for the actual problem is brought into question. In particular, if this is possible for a large number of examples (for example, different molecules) for the given use case, either the scope of the existing \textit{problem} needs to be revisited, or the possibility of quantum advantage is to be considered improbable. This demonstrates that benchmarking against optimized classical solvers, particularly those utilizing problem-specific approximations, often presents a significant obstacle to finding use cases for quantum computers. But at the same time finding a classical solution to achieve a satisfactory solution has obviouly also a substantial value for any end user. 

On a more possitive note for quantum advantage, the possibility to find broadly applicable use cases for quantum computing can be substanially enhanced by investing efforts of transforming the original industry problem into a formulation appropriate for the quantum computer. As an example, solving differential equations is probably one of the key promising problems in applications of quantum computing. It has been shown that in principle we expect quantum computers to be able to solve differential equation with an exponential speedup~\cite{harrow2009quantum,berry2014high,kiani2022quantum}. But, a differential equation is never the \textit{actual} industry problem. Usually, the actual problem requires the solution of a differential equation. However, at this point only a small amount of work is published on the question how to transform the actual problem onto the quantum computer without losing the expected speedup. To the best of our knowledge, reference~\cite{WilhelmMauch2024QUASIM} is among the only studies addressing this topic directly in the context of manufacturing engineering and some work on problem-specific differential equations for machine learning~\cite{Liu2024,wang2025efficientquantumalgorithms}. And much more should be possible.

In Table~\ref{tab:qa-criteria} we introduce our rating for these four ITBQ criteria and provide further details how to rank the use cases to track the process towards demonstrating quantum advantage. This table can be understood as a checklist and we will use the rating when describing the use cases in the following chapters. After discussing the two contrasting archetypes -- Industrial Optimization and Abstract Spin Problems -- in section~\ref{sec:two-contrasting-archetypes}, we will discuss three problems at different levels of readiness to demonstrate quantum advantage: Nuclear Magnetic Resonance in sections~\ref{sec:nmr}, Multireference Chemistry in section~\ref{sec:active-space-methods} and Radicals in section~\ref{sec:radicals}. Finally, we summarize our assessment for the discussed problems and conclude on the interpretation in section~\ref{sec:conclusions}.

\begin{table*}[htbp]
\centering
\renewcommand{\arraystretch}{1.2}
\rowcolors{2}{gray!15}{white}
\begin{tabular}{
    @{\hspace{0.15cm}}p{3.2cm}@{\hspace{0.15cm}}|
    @{\hspace{0.15cm}}p{3.2cm}@{\hspace{0.15cm}}|
    @{\hspace{0.15cm}}p{3.2cm}@{\hspace{0.15cm}}|
    @{\hspace{0.15cm}}p{3.2cm}@{\hspace{0.15cm}}|
    @{\hspace{0.15cm}}p{3.2cm}@{\hspace{0.15cm}}
}
\rowcolor{gray!30}
\textbf{Criteria}
    & \centering \starzero
    & \centering \starone
    & \centering \starthree
    & \centering \starfive
    \tabularnewline \hline
\raggedright\textcolor{hqstext}{\textbf{Identify Industry Problem}}
    & no clear problem, only area/field
    & academic problem
    & story relating to industry
    & proven pain point
    \tabularnewline
\raggedright\textcolor{hqstext}{\textbf{Transform to Quantum}}
    & no specification how to get input for QC
    & idea what is needed
    & prototype validated with real data
    & runs routinely for all relevant problems
    \tabularnewline
\raggedright\textcolor{hqstext}{\textbf{Get the Job Done Without the Quantum Computer}}
    & no idea what is a reasonable comparison
    & current solution in industry known
    & best (academic) solution known
    & best conventional solver available for comparison
    \tabularnewline
\raggedright\textcolor{hqstext}{\textbf{Show Quantum Advantage}}
    & no quantum algorithm identified
    & \raggedright QC algorithm identified
    & explanation why QA possible with chosen algorithm
    & implementation with clear QA reasoning
    \tabularnewline
\end{tabular}
\caption{Criteria to show quantum advantage -- \textbf{I}dentify, \textbf{T}ransform, \textbf{B}enchmark, Show \textbf{Q}uantum Advantage (ITBQ) -- with increasing completeness from left to right.}
\label{tab:qa-criteria}
\end{table*}

\section{Two Contrasting Archetypes: Industrial Optimization and Abstract Spin Problems}
\label{sec:two-contrasting-archetypes}

As the field of quantum computing matures, considerable attention has turned to the quest for “quantum advantage” in real-world applications. Two especially prominent families of use cases are the solution of industry-critical optimization problems and the simulation of abstract spin models, each highlighting key obstacles and open questions. Examining these two archetypes sharpens our understanding of both the potential and the current limitations in the search for valuable quantum applications.

\subsection{Industrial Optimization: High Impact, Unproven Quantum Edge}

Optimization problems underpin a vast array of real-world industrial processes. Mapping these problems onto quantum hardware is typically performed via transformation into the Quadratic Unconstrained Binary Optimization (QUBO) form—a mathematical structure well-suited to qubit-based quantum devices. However, this crucial \textit{Transform to Quantum} step is not straightforward in practice. Industry problems are often high-dimensional and involve diverse constraints, which means they are neither quadratic nor unconstrained. Thus, the process of converting them to standardized QUBO instances is not straightforward and rarely automated. Consequently, robust and user-friendly software for this transformation is still in its infancy. Most transformations are tailored case-by-case, limiting the broader applicability of quantum optimization techniques.

Additionally, there has been only a little emphasis on decomposing large, industry-scale optimization tasks so that only the truly hard or quantum-suitable subproblems are offloaded to a quantum computer, leaving the remainder to classical computation. In the most current workflows, the entire industry problem is mapped to a quantum formulation -- a strategy that may neither be optimal nor necessary. The lack of systematic framework or integrated software tools to identify and carve out quantum-suitable subproblems remains a significant obstacle.

Once transformed to a QUBO formulation, these problems are usually benchmarked against state-of-the-art classical solvers such as QUBO-specific solutions or general tools like Gurobi. While Gurobi and similar tools are highly effective, there is an ongoing argument that these classical baselines may themselves be open to further improvement, particularly by adapting classical heuristics or algorithms to the specific problem structures encountered in practice. This suggests that even the best “classical versus quantum” comparisons must be interpreted with caution -- continuous advances in conventional software could narrow or erase any incipient quantum advantage.

The overall landscape, as highlighted in recent benchmarking studies~\cite{Preskill2018nisq}, is that while industrial optimization problems are of truly exceptional practical relevance, current approaches of transforming them to quantum and benchmarking for quantum advantage remain immature~\cite{Costa2024quantum,Kikuura2024benchmarking,Hauke2024annealing}. Neither a uniformally established methodology exists to fairly demonstrate quantum edge on industrially meaningful optimization tasks, nor is there full maturity in the quantum-classical software toolchain that would make real-world adoption straightforward.

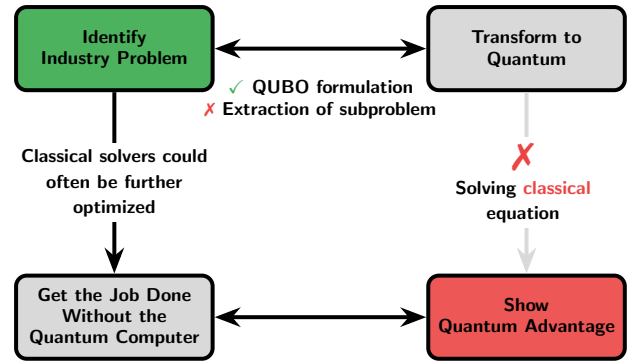
\begin{figure}[!b]
    \centering
    \begin{tikzpicture}[>=Stealth, node distance=2.8cm, font=\sffamily\scriptsize\bfseries, line width=1.5pt]
    
    \node (industry) [draw, rounded corners, align=center, minimum width=2.6cm, minimum height=1.1cm, fill=hqsgreen!95, text=black] {Identify \\ Industry Problem};
    \node (transform) [draw, rounded corners, align=center, right=of industry, minimum width=2.6cm, minimum height=1.1cm, fill=gray!30, text=black] {Transform to \\ Quantum};
    
    \draw[<->, thick, line width=1.5pt] (industry) -- (transform)
    node[midway, inner sep=1.2pt, below=10pt] {%
      \scriptsize
      \parbox[t]{3.1cm}{
        \textcolor{hqsgreen}{\checkmark} QUBO formulation \\
        \textcolor{hqsred}{\xmark} Extraction of subproblem
      }
    };  
 
    \node (classical) [draw, rounded corners, align=center, below=2.4cm of industry, minimum width=2.6cm, minimum height=1.1cm, fill=gray!30,text=black] {Get the Job Done\\ Without the\\ Quantum Computer};
    \node (quantum) [draw=black, rounded corners, align=center, below=2.4cm of transform, minimum width=2.6cm, minimum height=1.1cm, fill=hqsred!90,text=black] {\textcolor{black}{Show} \\ \textcolor{black}{Quantum Advantage}};
    
    \draw[<->, thick, line width=1.5pt] (classical) -- (quantum)
        node[midway, fill=white, inner sep=1.5pt, below=10pt] {};
       
    \draw[->, thick,line width=1.5pt] (industry) -- (classical)
    node[midway, fill=white, inner sep=2pt,] {%
      \scriptsize
      \parbox[t]{2.6cm}{
        \setstretch{1.2} 
          Classical solvers could often be further optimized
      }
    };  
 
      \draw[->, thick, line width=1.5pt, gray!30] (transform) -- (quantum)
      node[midway, fill=white, text=black, align=center, inner sep=2pt] {%
      \scriptsize
      \parbox[t]{2.6cm}{
        \setstretch{1.2} 
        \textcolor{hqsred}{\Large\xmark} \\ Solving \textcolor{hqsred}{\textit{classical}} equation
      }
    };  
    \end{tikzpicture}
  \caption{Evaluation of Industrial Optimization.}
  \label{fig:optimization}
\end{figure}

Despite the significant interest in applying quantum computing to industrial optimization problems, current use case studies reveal a lack of provable quantum advantage, whether exponential or polynomial. While it is anticipated that quantum algorithms may offer a polynomial speedup, this may prove insufficient given the slow gate times associated with error-corrected quantum computers~\cite{Babbush2021Focus}. Recent research on the Optimal Polynomial Intersection (OPI) problem provides at least some evidence, that exponential speedup could be achievable for finding approximate solutions~\cite{Jordan2024DQI}. However, it is important to note that the OPI problem does not directly relate to any of the practical optimization challenges currently under consideration. Thus, while the potential for quantum advantage exists, it remains largely unproven and disconnected from real-world applications.

\subsection{Abstract Spin Models: Plausible Advantage, Unclear Use}

Abstract spin models -- most notably the Ising model and its quantum and open-system generalizations -- hold a central role in the search for quantum advantage. These models are not only conceptually clean and naturally suited to quantum computer hardware, but also have been the subject of landmark experimental~\cite{king2023quantum, king2024supremacy} and theoretical milestones. Recent notable examples include IBM’s experimental quantum simulation of the kicked Ising model~\cite{PhysRevResearch.6.013326} -- a paradigmatic problem in quantum chaos, an experiment that was beyond the limit of brute force simulation although still classically solvable with the right approximative methods~\cite{PhysRevResearch.6.013326, Schollwock2011}. Another line of research, led by Google, has focused on bath-assisted Ising systems~\cite{Mi_2024}, where quantum dynamics are simulated in the presence of engineered environmental coupling (`bath”), further expanding the frontiers of classically challenging quantum simulation.

The belief in quantum advantage for these models rests on solid ground: for spin systems the Hilbert space scales like $2^N$ with $N$ being the number of spins. This exponential scaling with system size makes \textit{classical} exact simulations of quantum many-body spin dynamics rapidly intractable. Quantum computers can natively implement dynamics governed by these models and should be able to outperform classical simulation methods as system size increases. Theoretical bounds and early experimental data suggest that even for highly simplified spin systems, no known classical algorithm can reproduce their time-dependent behavior efficiently as quantum system size grows. This has also been rigourously proven for models as simple as the translationally invariant Ising Model~\cite{PhysRevLett.118.040502}.

However, the value of these results outside the quantum computing research community is less clear. While kicked rotors and bath-assisted Ising models are significant scientific testbeds, their known direct application to industrial or technological problems is minimal. They are abstract representatives of large classes of complex systems, but do not themselves encode the structure, constraints, or application relevance of real-world challenges. Thus, the added value of demonstrating quantum advantage with these problems is presently observed in benchmarking, hardware validation, and the demonstration of new paradigms in quantum simulation.

Ultimately, connection to important real-world problems is critical for these quantum advances to transcend their demonstration status. The scientific community is actively seeking representations of industry-relevant problems with these spin models. However, in the absence of direct practical benefit, these experiments, while impressive, do \textit{not} yet fulfill the promise of quantum advantage. 

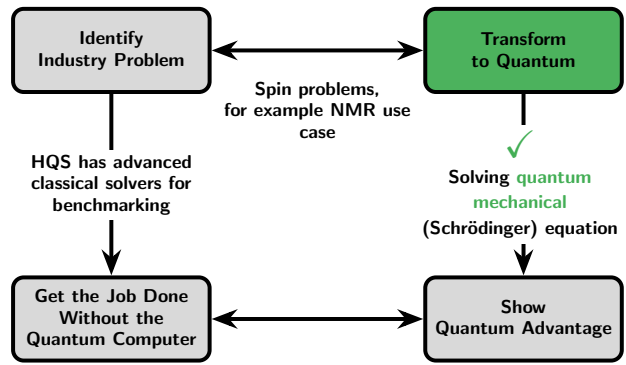
\begin{figure}[!tb]
    \centering
    \begin{tikzpicture}[>=Stealth, node distance=2.8cm, font=\sffamily\scriptsize\bfseries, line width=1.5pt]
    
      \node (industry) [draw, rounded corners, align=center, minimum width=2.6cm, minimum height=1.1cm, fill=gray!30, text=black] {Identify \\ Industry Problem};
      \node (transform) [draw, rounded corners, align=center, right=of industry, minimum width=2.6cm, minimum height=1.1cm, fill=hqsgreen!95, text=black] {Transform \\ to Quantum};
     
      \draw[<->, thick, line width=1.5pt] (industry) -- (transform)
      node[midway, inner sep=1.2pt, below=10pt] {%
        \scriptsize
        \parbox[t]{2.8cm}{
          Spin problems, \\ for example NMR use case
        }
      };  
    
      \node (classical) [draw, rounded corners, align=center, below=2.4cm of industry, minimum width=2.6cm, minimum height=1.1cm, fill=gray!30,text=black] {Get the Job Done\\ Without the\\ Quantum Computer};
      \node (quantum) [draw, rounded corners, align=center, below=2.4cm of transform, minimum width=2.6cm, minimum height=1.1cm, fill=gray!30,text=black] {Show \\ Quantum Advantage};
     
      \draw[<->, thick, line width=1.5pt] (classical) -- (quantum)
          node[midway, fill=white, inner sep=1.5pt, below=10pt] {};
         
      \draw[->, thick,line width=1.5pt] (industry) -- (classical)
      node[midway, fill=white, inner sep=2pt,] {%
      \scriptsize
      \parbox[t]{2.6cm}{
      HQS has advanced classical solvers for benchmarking
      }
      };  
    
        \draw[->, thick, line width=1.5pt] (transform) -- (quantum)
        node[midway, fill=white, text=black, align=center, inner sep=2pt] {%
        \scriptsize
        \parbox[t]{2.6cm}{
          \setstretch{1.2} 
          \textcolor{hqsgreen}{\Large\checkmark} \\ Solving \textcolor{hqsgreen}{\textit{quantum mechanical}} (Schrödinger) equation
        }
      };  
      \end{tikzpicture}
  \caption{Evaluation of Abstract Spin Models.}
  \label{fig:abstractspinmodels}
\end{figure}

\subsection{Evaluation and the Ongoing Challenge}

These two archetypes encapsulate the state of quantum computing applications today. On the one hand, we have optimization problems with immediate and enormous industrial impact, but for which convincing evidence of quantum advantage remains elusive (Fig.~\ref{fig:optimization}). Although, the problem can be transformed to quantum and formulated in form of a Hamiltonian, the equation solved is a \textit{classical} minimization or maximization problem, and not a quantum mechanical equation which might naturally benefit from a quantum computer. Along the way, the available classical solvers for optimization problems have the potential for further improvements making benchmarking to demonstrate quantum advantage ever more unattainable.

On the other hand, abstract spin problems present highly plausible opportunities for quantum speedup (Fig.~\ref{fig:abstractspinmodels}). The formulation of a spin system is naturally suited to be time evolved by the means of solving the \textit{quantum mechanical} Schrödinger equation. These is the reasons why we pursue their solution and also invest efforts into continuous improvement of equivalent classical solvers. To connect these abstract spin problems to actual use cases, the most important step is \textit{transform to quantum}, which we will discuss in section \ref{sec:nmr}.  However, the ultimate practical value of abstract spin problems, beyond the NMR use case, stays ambiguous. 

Both archetypes highlight the centrality of rigorous problem selection, fair benchmarking, and honest appraisal of what quantum computing can, and cannot (yet), do.

\begin{figure*}[!htbp]
  \centering
  \includegraphics[width=0.9\textwidth]{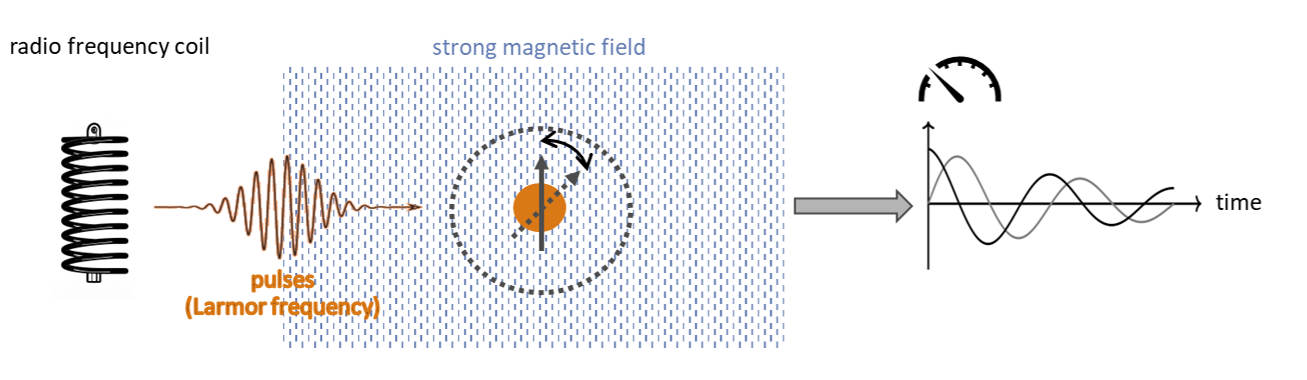}
  \caption{Principles of an NMR spectrometer: the sample is placed in a strong magnetic field, causing a net magnetization of the nuclear spins by the static field. Pulses from radio frequency coils, in resonance with the Larmor frequency of the nuclei, cause an oscillating transverse magnetization. Measurement of the decaying response after completion of the pulse sequence leads to the NMR spectrum after processing.}
  \label{fig:nmrspectrometer}
\end{figure*}

\section{Use cases in quantum simulation}

Quantum chemistry represents a promising frontier for the application of quantum computing, as it seeks to address complex problems that are inherently quantum mechanical in nature. In this section, we will explore three significant use cases: Nuclear Magnetic Resonance (NMR), multireference quantum chemistry, and the study of radicals. Each of these use cases will be examined through the lens of the four critical ITBQ criteria for evaluating quantum applications: identifying the industry problem, transforming the problem into a quantum format, assessing whether classical methods can suffice, and demonstrating the potential for quantum advantage.

A substantial effort at HQS has been dedicated to the  \textit{transform to quantum} step for these use cases, enabling us to extract the most relevant quantum mechanical degrees of freedom. By systematically analyzing these use cases, we aim to highlight the unique challenges and opportunities presented by quantum computing in the realm of quantum chemistry, ultimately shedding light on the potential for achieving significant advancements in this field. But also show a more fundamental approach about how to judge use cases and guide more projects towards quantum advantage.

\subsection{Nuclear Magnetic Resonance}
\label{sec:nmr}

\subsubsection{Identify industry problem}

Nuclear magnetic resonance (NMR) spectroscopy is one of the most important analytical techniques in chemistry. When determining or elucidating the structure of organic compounds, it is usually NMR spectroscopy that provides the majority of information, complemented by other techniques such as infrared spectroscopy or mass spectrometry. The information content of NMR spectra is only superseded by X-ray crystallography. However, NMR spectra are usually recorded for substances in solution without
needing to prepare crystals, rendering it much more suitable as a routine technique. In contrast to crystallographic structure determination, which often requires highly specialized training, interpretation of NMR spectra is considered a core skill and is introduced early in undergraduate chemistry courses~\cite{Clayden2012}. Whenever organic (and many inorganic) substances are synthesized in a modern lab, it is likely that NMR spectra are recorded, too.

In an NMR spectrometer (see~Fig. \ref{fig:nmrspectrometer}), the sample is placed in a strong magnetic field (\qty{9.4}{\tesla} in a routine \qty{400}{\mega\hertz} spectrometer), causing a net magnetization of the nuclear spins by the static field. Pulses from radio frequency coils, in resonance with the Larmor frequency of the nuclei, cause an oscillating transverse magnetization; measurement of the decaying response after completion of the pulse sequence leads to the NMR spectrum after processing. Thus, NMR problems can be modelled through the time evolution of a system of nuclear spins, interacting with the external field, the oscillating pulses and among each other~\cite{Levitt2008}.

The interaction with the molecule hosting the nuclei is mainly absorbed into two types of parameters:
\begin{enumerate}
  \item Nuclei in a molecule are diamagnetically shielded, leading to an energy splitting that is slightly different than that of a bare nucleus in the same magnetic field.\footnote{NMR spectroscopy is usually performed
  with diamagnetic samples. NMR spectroscopy of paramagnetic compounds also exists; spectra are profoundly different from those of diamagnetic samples.} Moreover, shielding experienced by nuclei differs within the same molecule, being influenced by functional groups, adjacent substituents and their relative arrangement in space. This effect is quantified through \emph{chemical shift} values. Indeed, it is the chemical shifts that permit NMR to be used as an analytical tool for structure determination in the first place.
  \item Observable interactions between spins are due to indirect dipole-dipole couplings (\emph{$J$-couplings}) between spins, which are mediated by chemical bonds. These couplings are specific for the type and number of bonds separating nuclei, thus providing valuable structural information.
\end{enumerate}

A particularly common variant of NMR spectroscopy is $^1$H-NMR. Protons in
molecules form systems of interacting spins with different chemical shifts, thus leading to a particularly complicated appearance of spectra.
In practice, $^1$H spectra are usually assigned by identifying chemical shifts according to their characteristic values, and by analyzing coupling patterns, where possible.

Alternatively, analysis of an NMR experiment can proceed via its quantum mechanical simulation. Recent opinions in chemical and pharmaceutical publications argue in favor of putting more emphasis on the rigorous simulation approach \cite{Botros2025,Achanta2021,Zhang2024}. The size of the exact representation of the spin system scales exponentially with the number of spins. In order to set up the problem description, values of the NMR parameters---chemical shifts and $J$-couplings---are needed. These can either be fitted to the spectrum, typically in an iterative procedure requiring repeated spectrum simulation. Alternatively, the NMR parameters can be predicted, likely with some degree of deviation from experiment. Predicted values can also
be used as a starting point for further refinement, or be combined with experimental parameters to substitute for missing values.

\begin{figure}[!bt]
  \centering
  \includegraphics[width=0.95\columnwidth]{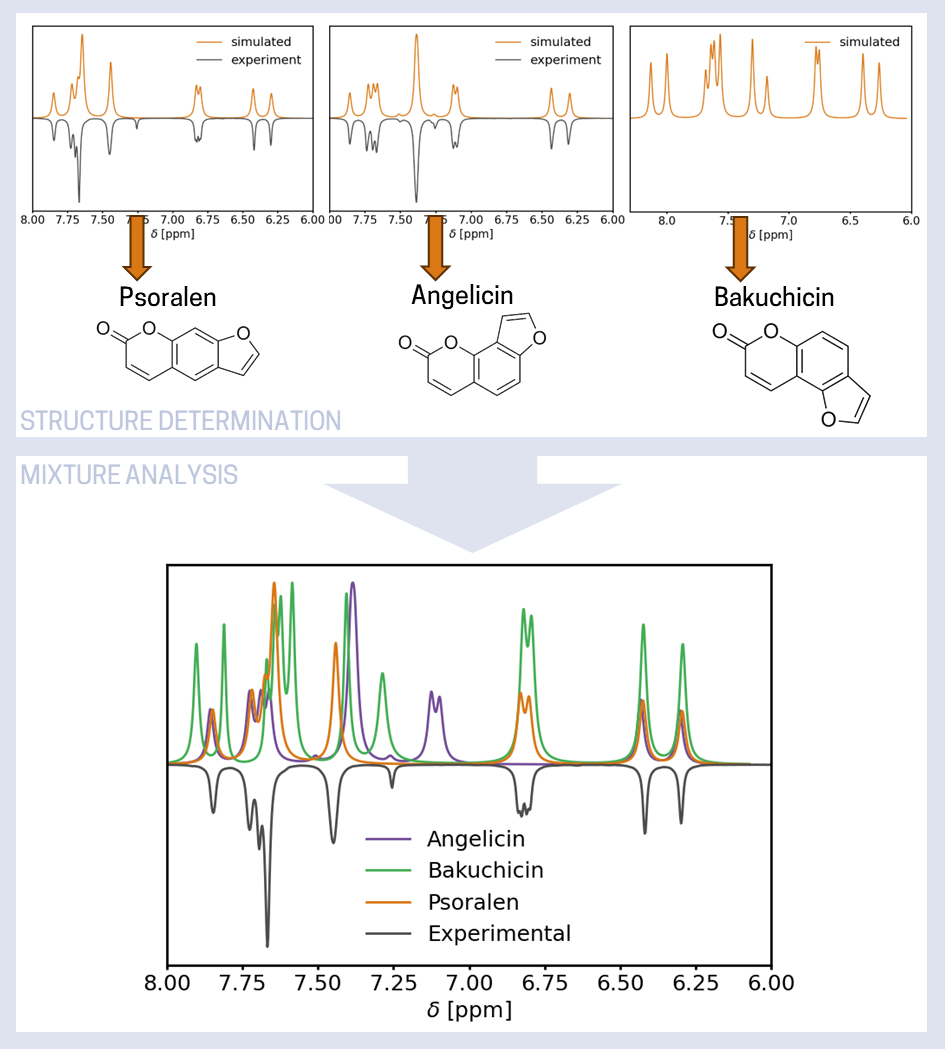}
  \caption{Two NMR applications -- structure determination and mixture analysis -- are combined in a single workflow. Experimental spectra for Psoralen and Angelicin were provided by PhytoLab GmbH \& Co. KG, while no experimental data were available for Bakuchicin; its spectrum is simulated only. Although all three molecules share the same molecular formula ($\mathrm{C}_{11}\mathrm{H}_6\mathrm{O}_3$), their spectra are distinct, highlighting the necessity for precise NMR analysis.}
  \label{fig:nmrapplications}
\end{figure}

One important NMR application is \textbf{structure determination} (see~Fig.~\ref{fig:nmrapplications}~(upper panel)), which is highly relevant both in academic and in industrial research: the structure of a molecule is unknown or uncertain. To elucidate the structure, it is necessary to assign NMR parameters such that they reproduce the experimental spectrum; at the same time, the values of the NMR parameters need to be entirely consistent with, and not in contradiction to, the features of the proposed molecular structure.
Moreover, there should be no other plausible structures that could also be in agreement with the experimental spectrum.
To aid with structure determination, spectrum simulation can be coupled with computational NMR parameter prediction for proposed chemical structures.

Another important NMR application is \textbf{analysis of substance mixtures} (see~Fig.~\ref{fig:nmrapplications}~(lower panel)), for example through quantitative NMR (qNMR). Important application areas include quality control, such as the purity determination of pharmaceuticals; analysis of biofluids, for example in metabolomics; or real-time monitoring of chemical or biological processes~\cite{Giradeau2017,Giradeau2023}. In these cases, structures of target molecules are known in principle, and often also their reference spectra. The goal is to determine substance concentrations, or the presence of impurities. Concentration determination through NMR is based on the principle that integrals over peaks in the spectrum are proportional to the number of resonating nuclei. In these applications, NMR often serves as an alternative or as a complement to mass spectrometry coupled with liquid or gas chromatography (e.g., GC/MS, LC/MS).

Applying qNMR requires either spectra of reference compounds to be recorded, or libraries of reference spectra to be built, such as in the Human Metabolome Database~\cite{Wishart2021}. Spectra of compounds depend on the strength of the magnetic field specific to the spectrometer, but they can be recomputed if the NMR parameters are available. The NMR parameters themselves, especially the chemical shifts,
can be influenced by conditions such as solvent, temperature, concentration or pH value. For metabolomics, the GISSMO database contains sets of spin parameters that have been fitted to experimental spectra for hundreds of compounds~\cite{Dashti2017,Dashti2018}.

For the most NMR applications, typical NMR spectrometers employ superconducting magnets to maintain high magnetic fields. Benchtop devices with lower fields generated by permanent magnets are becoming increasingly popular as a less expensive and more compact alternative~\cite{Grootveld2019}. With field strengths of ca.~\qtyrange{1}{2}{\tesla}, the appearance of NMR spectra becomes profoundly more complicated, rendering accurate simulation even more useful for the analysis of such spectra.

Our assessment of NMR simulation as an actual industry problem is four out of five stars (\starfour). The importance of nuclear magnetic resonance as an analytical technique is beyond doubt. Realistic spectra can be calculated, subject to having or computing suitable NMR parameters. This use case does not receive the highest rating, since additional efforts are required to achieve a seamless integration of simulations within the established experimental analytical workflow.

\subsubsection{Transform to quantum}

A central part in the description of the NMR problem is the Spin Hamiltonian~\cite{Levitt2008}:

\begin{equation}
  \hat{H} =
    -\sum_{k} \gamma_k \left( 1 + \delta_k \right) \mathbf{B} \hat{\mathbf{I}}_k
    + 2 \pi \sum_{k > l} J_{kl} \hat{\mathbf{I}}_k \hat{\mathbf{I}}_l .
\end{equation}

In this equation, $\mathbf{B}$ is the static magnetic field; the operator
$\hat{\mathbf{I}}_k = \hbar^{-1} \hat{\mathbf{S}}_k$ represents the spin of nucleus $k$, and
$\gamma_k$ is the gyromagnetic ratio of the respective isotope in vacuum. Chemical shifts
$\delta_k$ quantify the diamagnetic shielding of each nucleus within the molecule.
Values of chemical shifts are specified in parts per million (ppm), as the diamagnetic shielding is very weak: for example, chemical shifts of $^1$H range within \qty{30}{ppm} (the majority in a range of \qty{12}{ppm}), those of $^{13}$C within \qty{250}{ppm}. Indirect dipole-dipole coupling constants $J_{kl}$ quantify the interaction between nuclei mediated through chemical bonds. Magnitudes of $J$-coupling constants between pairs of protons are typically in a range of up to \qty{20}{\hertz}, though values of several hundred \unit{\hertz} occur between nuclei separated by only one bond.

The Spin Hamiltonian presented above is sufficient for spin-$\frac{1}{2}$ nuclei of diamagnetic molecules in liquid state. Direct dipole interactions between nuclei cancel out due to molecular rotations occurring on a faster timescale than the NMR experiment, though they become important for macromolecules or solids.

End-to-end NMR simulation, such as for structure elucidation, requires computational prediction of chemical shifts $\delta_k$ and $J$-couplings $J_{kl}$ for arbitrary molecular structures. A fairly general procedure can be set up with the help of quantum chemistry, as exemplified by a workflows deployed at HQS Quantum Simulations for liquid-state NMR:

\begin{enumerate}
  \item Input of a molecule is provided most often as a 2D structural formula, which needs to be converted to a three-dimensional geometry. To achieve this, it was necessary to build a a driver engine for the open-source packages RDKit and OpenBabel~\cite{RDKit,OBoyle2011} mixed with fast consitency checks.
  \item Many molecules exist in an equilibrium between different conformational structures. These structures interconvert on a much faster timescale than that of the NMR experiment. Therefore, the NMR parameters need to be averaged properly over all thermodynamically accessible conformers. We use a procedure that combines the open-source software CREST to find relevant structures automatically with additional methods for further improvements of the structural ensemble~\cite{Pracht2024}.
  \item Geometries of relevant conformers are optimized with the help of density functional theory (DFT). Vibrational frequency calculations are performed to verify that the optimized geometries correspond to energy minima rather than saddle points, and to evaluate thermodynamic contributions to their Gibbs energies. Together with refined electronic energies, the Gibbs energies are used to select the most important conformer structures for the subsequent NMR parameter calculations. In this and in the next step, we use the ORCA software under a commercial license~\cite{Neese2022}.
  \item Shieldings and J-couplings are calculated with DFT for all relevant conformer geometries.
  \item The NMR parameters are averaged using the Gibbs energies of the individual structures. Shieldings are converted to chemical shifts.
\end{enumerate}

The steps above are a simplified representation of the actual procedure, which has been inspired by published academic work~\cite{Grimme2017}.

Besides direct computation of NMR parameters using quantum chemistry, various data-driven approaches have been developed over the course of decades, some of them integrated into widely used NMR analysis software. We refer the interested reader to a review that focuses on chemical shift prediction, but also discusses $J$-couplings~\cite{Jonas2022}. Traditionally, such approaches make use of data collections of experimental assignments, though some of the more recent approaches also incorporate computed data sets.

 Procedures to estimate NMR parameters either using quantum chemistry or using data-driven approaches are in principle well-established, but software implementations which perform consistently in an automated fashion are still rare. At the same time, computed NMR parameters cannot have a perfect agreement with experiment. In addition, some experimental factors, such as the impact of solvents or interactions between solute molecules, can be very difficult to model. NMR simulations need to be integrated either in existing NMR analysis solutions or into new software. Details of the integration depend on the specific use case. We refer to a perspective on computer-aided structure elucidation as an example~\cite{Elyashberg2021}. We assess our current state of our transformation and extraction to the quantum mechanical degress of freedom with four stars (\starfour), since we so far achieved a high level of automation and precision. However, even in our implementation, a noticeable number of molecules still need to be checked by hand at this stage, indicating that there is still room for improvement.

\subsubsection{Get the job done without the quantum computer}

In order to demonstrate quantum advantage, it is necessary to compare quantum simulations to their best available conventional counterparts. The dimensionality of the NMR problem in exact representation scales with the number of spins $N$ as $O\left(2^N\right)$. This means even to relatively small problems involving around $25$ spins, exact numerical solution becomes impossible. But there are still approximative solutions that allow a simulation of the problem~\cite{KUPROV2007241, KUPROV201131, EDWARDS2014107, KUPROV2016124, kuprovtutorial2018}. We primarily consider two conventional options: one is the HQS in-house NMR solver, that will be elaborated upon below. The other is the well-established open-source Spinach software~\cite{Hogben2011}.

\begin{figure}[!tbp]
  \centering
  \includegraphics[width=0.95\columnwidth]{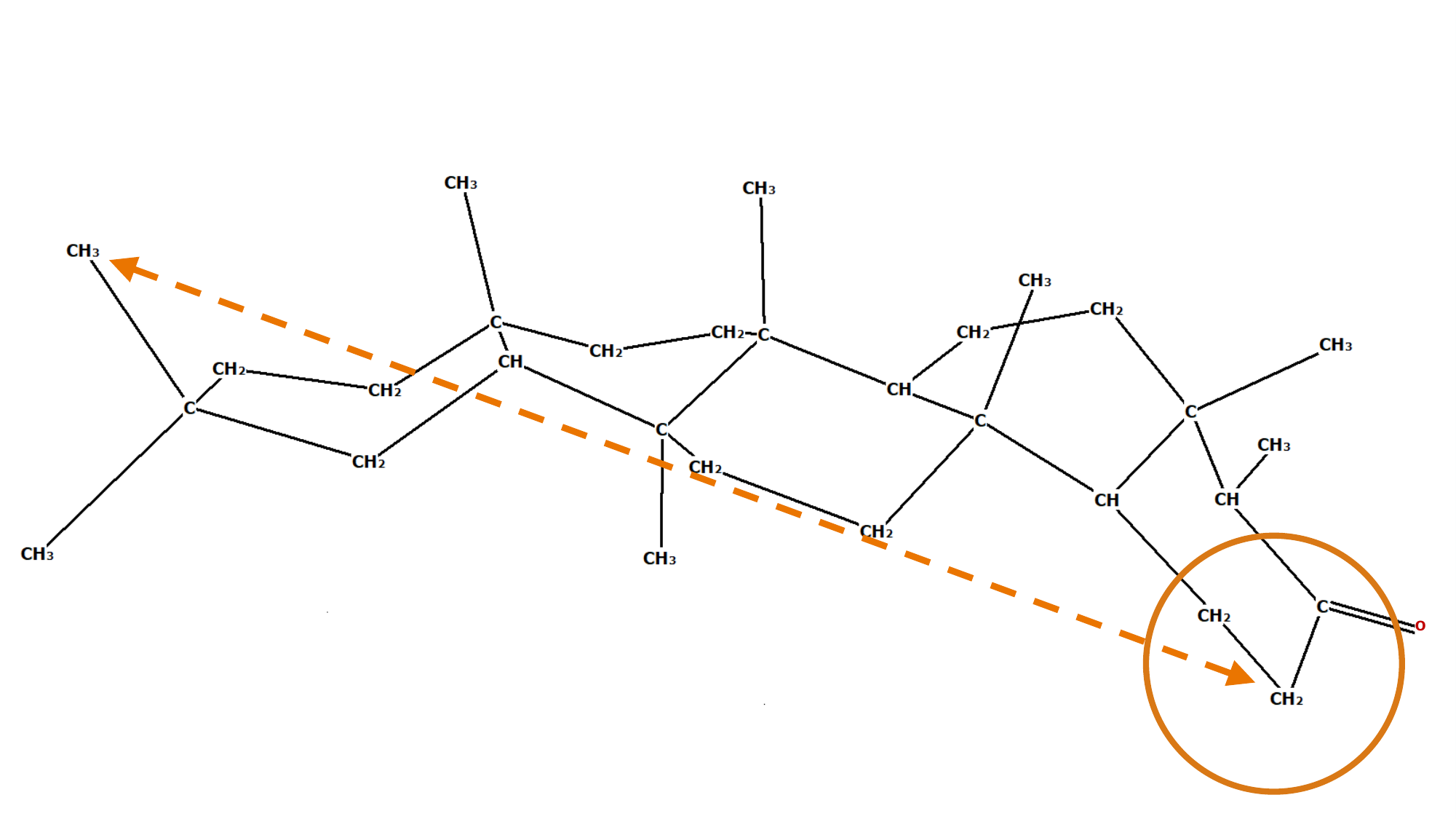}
  \caption{If we look at large molecules like Friedelin (depicted as a 2D structure), it is not necessary to simulate every spin interacting with every other spin. As shown by the arrow, distances between nuclear spin of hydrogen atoms can be large and then it is only necessary to simulate a local cluster as depicted by the circle.}
  \label{fig:clustering}
\end{figure}

For many molecules of practical relevance under high-field conditions, it is not necessary to calculate the NMR spectrum exactly---sufficiently accurate solutions can be obtained using suitable clustering approximations (see Fig.~\ref{fig:clustering}). This way, the HQS solver can calculate 1D spectra for organic molecules with tens of protons in a matter of minutes. Thus, it is well-suited to explore more challenging conditions: for example, under lower magnetic fields, where energy splittings decrease while couplings remain constant. Results in this parameter regime are shown in Fig.~(\ref{fig:connectivity}) and discussed in the next section.\\
In addition, symmetry considerations can be used to considerably speed up calculations, even though the exponential scaling cannot be eliminated. This allows even complicated systems to be simulated without the clustering approximation, such as 1,2-di-tert-butyl-diphosphane, which contains 22~spins: twenty $^1$H nuclei and two $^{31}P$ nuclei that form one coupled system (see Fig.~\ref{fig:diphosphane} and discussion in the next section).\\
The input and output of this simulation use case are unambiguously defined, and efficient conventional software is available. Therefore, we rate the benchmarking opportunity with five stars (\starfive).

\subsubsection{Show quantum advantage}

In this subsection, we will shortly review the quantum mechanical foundations of NMR simulation on a quantum computer to elucidate the potential arguments for quantum advantage. Simulating the time evolution of quantum systems on classical computers is computationally demanding, particularly for large and highly entangled systems. In NMR simulations, the quantity of interest is the time-dependent expectation value of the spin component in the direction perpendicular to the applied magnetic field. Assuming a strong magnetic field along the z-direction, spins precess in the transverse x-y-plane. We consider the total magnetization along the x-direction, written as \( I^x \).
The expectation value evolves over time according to
\begin{equation}
\langle I^x(t) \rangle = \langle I^x \rho(t) \rangle.
\end{equation}
Here, \( \rho(t) \) is the density matrix describing the system at time \( t \), and \( I^x \) represents the collective spin operator summing over all individual spins. The evolution of \( \rho(t) \) follows the Liouville-von Neumann equation:
\begin{equation}
\frac{d}{dt} \rho = -i [H, \rho].
\end{equation}
Since \( \rho \) is a \( 2^N \times 2^N \) matrix for an \( N \)-spin system, the memory and computational cost of an exact classical simulation grows as \( O(2^{2N}) \), making numerically exact large-scale simulations infeasible.

In contrast, quantum computers efficiently simulate NMR dynamics by evolving the system directly under unitary time evolution, avoiding the exponential memory cost of classical simulations. The time evolution follows \( \rho(t) = e^{-i H t} \rho(0) e^{i H t} \), enabling the study of larger spin systems. To derive the expression for the time-dependent expectation value of \( I^x \), we start with the system in thermal equilibrium, where the density matrix follows the Boltzmann distribution. In an NMR experiment, a \( \pi/2 \)-pulse applied along the x-axis rotates the magnetization from the z-direction to the y-direction, transforming the density matrix accordingly. This maps the total spin operator from the z- to the y-direction. At low temperatures, the density matrix can be approximated as proportional to the total spin operator along y. The system then evolves under the NMR Hamiltonian, and applying unitary time evolution to the transformed density matrix leads to the final expression:
\begin{equation}
\langle \hat{I}^x \rangle (t) \propto \langle \hat{I}^x e^{-i H t} \hat{I}^y e^{i H t} \rangle.
\end{equation}
To compute this expression on a quantum computer, the system is initially prepared in an eigenstate of \( \hat{I}^y \) with eigenvalue \( m^y_i \). This leads to the form~\cite{khedri2024impact}
\begin{equation}
\langle I^x(t) \rangle \propto \sum_{m^y_i} m^y_i \langle m^y_i | e^{iHt} I^x e^{-iHt} | m^y_i \rangle.
\end{equation}
Here, the sum runs over all initial states \( |m^y_i\rangle \), weighted by their respective eigenvalues \( m^y_i \). On a quantum computer, this calculation is performed by first preparing the initial quantum state encoding the eigenbasis of \( \hat{I}^y \), followed by evolving the system under the Hamiltonian using quantum simulation techniques such as Trotter decomposition. Finally, quantum measurements are carried out to extract the expectation values required for reconstructing the NMR spectrum.

\begin{figure}[t]
\centering
\includegraphics[width=1.0\columnwidth]{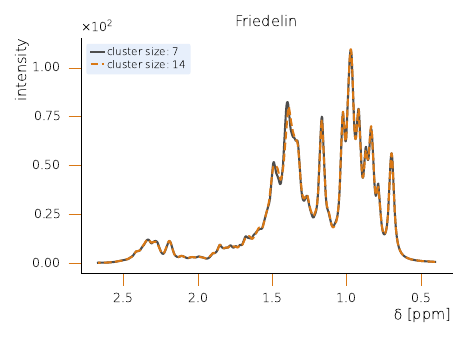}
\caption{NMR spectrum of Friedelin, a 50-spin molecule, simulated using a cluster-based solver approach. The molecule was divided into independent clusters, treated separately, with a broadening of 10 Hz. The simulation results demonstrate good convergence, with minimal changes in the spectrum as the cluster size increases.}
\label{fig:connectivity}
\end{figure}

The summation over initial states in the equation above scales exponentially with system size, making exact evaluation infeasible. Instead, quantum computing samples the result. This combination of efficient time evolution and sampling makes quantum computing useful for simulating NMR dynamics. It has been shown in previous work that sampling a polynomial number of initial states should be sufficent \cite{Sels2020}.

\begin{figure}[!tb]
\centering
\includegraphics[width=\columnwidth]{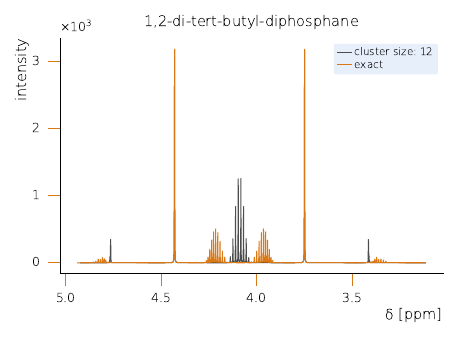}
\caption{NMR spectrum simulation at 500~\text{MHz} of 1,2-di-tert-butyl-diphosphane, a 22-spin molecule, comparing the results of a cluster-based approximation and an exact approach utilizing local SU(2) symmetry. The approximate spectrum (black) is obtained using a spin-dependent clustering method, which provides a computationally efficient solution but does not even qualitatively capture the relevant spin correlations. In contrast, the exact solution (orange) incorporates local SU(2) symmetry to reduce computational complexity while maintaining full accuracy.}
\label{fig:diphosphane}
\end{figure}

While quantum computing offers an exponential advantage for NMR simulations, classical algorithms have also been developed to efficiently treat large spin systems. One particularly effective approach is spin clustering (see Fig. (\ref{fig:clustering})), which reduces computational complexity by grouping strongly interacting spins together, enabling approximate solutions. In many cases, especially at high magnetic fields, this method can provide accurate results with cluster sizes of 8–12 spins, even for complex molecules.
For example, in the case of Friedelin (Fig.~\ref{fig:connectivity}), a 50-spin molecule, the cluster-based solver shows good convergence, with minimal spectral changes as the cluster size increases. This demonstrates that for certain systems with relatively weak spin correlations, spin clustering provides a feasible classical approach.

However, when strong correlations extend across the system, as in the case of 1,2-di-tert-butyl-diphosphane (Fig.~\ref{fig:diphosphane}), the limitations of spin clustering become apparent. This molecule consists of 22 coupled spins, where cluster approximations cannot fully capture the spin interactions. Instead, solving the system exactly requires the exploitation of local SU(2) symmetry, which significantly reduces the computational complexity while maintaining full accuracy. Without symmetry considerations, the problem would require the diagonalization of a $2^{22} \times 2^{22}$-dimensional matrix, making classical exact diagonalization basically intractable. Given examples like 1,2-di-tert-butyl-diphosphane, it seams likely that parameter examples and parameter regimes exist in which classical solutions remain challenging. At the same time it is clear that classical solvers can cover a large part of the relevant cases for NMR. Therefore we rate the probability to achieve quantum advantage with four stars (\starfour).

\subsection{Multireference Chemistry}
\label{sec:active-space-methods}

\subsubsection{Identify industry problem}

This section assesses Multireference Chemistry for ground and excited electronic states as a quantum
computing use case. The active space finder (ASF) software~\cite{ASF} is highlighted as a helpful
utility for preparation steps.

An exact solution to the electronic Schrödinger equation (in a given atomic basis set) can be
calculated with the full configuration interaction (FCI) method. Due to the exponential scaling of
its computational cost with system size, solving the FCI problem is only feasible for very small
systems. Being able to map the problem onto the states of a quantum computer appears to make
electronic structure an attractive target for potential quantum computing applications~\cite{Bauer2020,Abrams1999}.

Practically useful conventional solutions to the electronic structure problem invoke
approximations. A very popular and versatile one is density functional theory (DFT), which has been a fixture in quantum 
chemistry during the past decades and will likely remain so in the foreseeable future~\cite{Bursch2022}.
Single-reference electron correlation methods constitute a family of approaches that
systematically improves upon a single Slater determinant, typically from a Hartree-Fock
calculation~\cite{SzaboOstlund,PurpleBook}.

One of the comparably most affordable methods that performs significantly better for ground states
than DFT is the coupled cluster method with single and double excitations, and a perturbative
treatment of triple excitations (abbreviated as CCSD(T))~\cite{Bartlett2007}.
The computational cost of CCSD(T) has a formal scaling of $O\left(N^7\right)$ with the
system size $N$, restricting it to molecules with relatively few atoms in practical applications.
Local correlation techniques introduce approximations to single-reference methods that reduce their
steep scaling. They have matured to a point where local CCSD(T) calculations can be performed
routinely to calculate single-point energies of systems with tens or even hundreds of
atoms~\cite{RiplingerCCSD2013,RiplingerTrip2013,Ma2018,Nagy2024}.
On the other end of the spectrum, semi-empirical electronic structure methods are popular for
screening purposes or to simulate large systems~\cite{Bannwarth2020}.

Conceptually, quantum chemistry divides electron correlation into two types: dynamic correlation
is always present in many-electron systems, and amounts to many relatively small contributions by
all electrons. Static correlation occurs when a single Slater determinant is a qualitatively poor
approximation to the wave function, typically as a consequence of (near-)degenerate frontier orbitals. Such systems are commonly treated with multi-reference methods. Among these,
complete active space (CAS) methods are common representatives, particularly the complete active space self-consistent field (CASSCF) and complete active space configuration interaction (CASCI) methods~\cite{Olsen2011,Roos2016}.
It is usually possible to identify a subset of orbitals hosting
strongly correlated electrons, referred to as the active space. Within this subset, a full configuration interaction calculation is performed. The remaining orbitals, and their interaction
with the active space, is treated at a mean-field level.

The straightforward partitioning of the system in active space methods has made them an attractive target
for the quantum computing community: the active space would be mapped on the quantum computer,
while the rest of the system is treated conventionally. However, active space methods have their
own shortcomings. One of them is the need to select an appropriate active space. Another
difficulty is the adequate treatment of dynamic correlation of all electrons (in both active and
inactive orbitals), which is usually needed for quantitatively satisfactory results.

Electronic structure calculations as an industry use case (with or without quantum computing) are tied to problems that can be solved with quantum chemistry in the first place: something that needs
to be assessed on a case-by-case basis.
We refer to ref.~\cite{Deglmann2014} for a discussion of examples with industrial relevance. 
This process necessarily involves model building to capture the essential physics and chemistry of the target system. For instance, a comparably simple system to simulate is a molecule in gas phase. Calculating the impact of a solvent is already more complicated: very often implicit solvation models are used, which surround the solute with a dielectric continuum to avoid
simulating an ensemble of explicit solvent molecules. Setting up models of catalyst surfaces or of
protein sites can be profoundly complicated tasks. An example
is ref.~\cite{Drosou2024}, describing a combined multi-reference and multi-scale approach for a
photosynthetic reaction center.
Significant sources of quantitative errors in quantum chemistry include the computation of
thermodynamic corrections~\cite{Grimme2012}, and relativistic effects become important for heavy
elements. Thus, other aspects can be equally important as the electronic structure method itself.

When deciding whether to use an active space method, it is advisable to determine if strong
correlation is relevant for the system at hand. This does not appear to be the case
for the majority of problems in molecular ground state chemistry, including many
transition metal coordination compounds~\cite{Izsak2023,Shee2021}. In such cases, employing active
space methods is at best unnecessary and at worst counterproductive. Especially for transition
metal compounds, it is not uncommon to find debates about the need to employ multi-reference
methods in specific cases~\cite{Feldt2022,Drosou2022}.

Some transition metal systems possess pronounced multi-reference character. Notorious
examples include antiferromagnetically compounds with multiple metal centers.
Applications of multi-reference methods towards catalytic model systems have been reviewed in
ref.~\cite{Gaggioli2019}. For a systematic analysis with guidelines, we refer to
ref.~\cite{Shee2021,Neugebauer2023}.

Multiconfigurational character is encountered more commonly for excited electronic states than
for ground states. Therefore, active space methods can be relevant to model electronic excitations
in spectroscopy or in photodynamic processes~\cite{Lischka2018,Mai2020,Gonzalez2011}.

Finally, a sufficiently large active space is needed to demonstrate quantum
advantage over a conventional calculation. The number of orbitals in an active space calculation
should be appropriate for the system at hand: neither too large, nor too small.
Sometimes, one encounters a misconception which assumes that the quality of an active space
calculation generally increases with growing size of the active space.
This would be the case if the calculation was most appropriately described as truncated full
CI. However, active space calculations are normally performed only with the most important
orbitals, without recovering large parts of the dynamic correlation energy.
In this regime, one cannot expect meaningful convergence of computed properties, such as energy
differences, towards the exact result by extending the active space.

To summarize, a use case would involve a statically correlated system that requires a large active
space for its qualitatively correct treatment. Potential interesting examples are covered in review
articles on the application of multi-reference methods towards excited electronic
states~\cite{Lischka2018,Mai2020,Gonzalez2011} and on transition metal systems~\cite{Khedkar2021}.
To conclude the assessment, we rate the relevance of this use case with three stars
(\starthree) -- partly in light of the academic interest in the topic. 

\subsubsection{Transform to quantum}

A starting point for any quantum chemical electronic structure calculation is the definition of
an appropriate three-dimensional molecular geometry. A valid, though laborious approach is to build
it manually using a graphical three-dimensional structure editor.
For many molecules it is possible to convert a ``two-dimensional'' molecular connectivity
representation, such as a SMILES string or a Molfile, automatically to a three-dimensional
structure. This works well for typical organic molecules, but may be more challenging for
structures with metals or with unusual valences. Depending on the molecule, it may be advisable to
perform a conformer search. For various solids, it may be possible to obtain a suitable geometry from a crystallographic
database. In addition to the molecular geometry, one also needs to specify the
charge and spin multiplicity state of the system.

To determine a suitable geometry, such as an equilibrium structure, it is usually
necessary to perform a geometry optimization. The most rigorous approach would be a geometry
optimization with the respective multi-reference method. Often,
the multi-reference method is used only for a single-point energy calculation, based on
a geometry that was optimized beforehand with a less sophisticated approach, such as DFT. It should
also be noted that the geometry optimization and conformer search described in section~\ref{sec:nmr}
is mostly suited for single reference molecules which are of major interest in NMR.

Having chosen a suitable molecular setup, the active space needs to be defined. This is generally a
nontrivial task. Only the most strongly correlated orbitals are part of the active space. Unless
the molecule is very small, the majority of orbitals remain inactive. Therefore, the selected active
space has a large influence on the outcome of the calculation. There are no definitive, rigorous
criteria for active orbital selection---usually it is based on experience and chemical
intuition~\cite{Veryazov2011}. This is an undesirable situation, as it introduces subjectivity and makes such calculations very
difficult to integrate into automatic workflows. Several schemes have been published that aim to
put active space selection onto a more systematic footing~\cite{Pulay1988,Stein2016,Sayfutyarova2017,Khedkar2019,Bensberg2023}

At HQS Quantum Simulations, the ``Active Space Finder'' software (ASF) was developed in collaboration with Covestro to address
this problem~\cite{ASF}. The starting point is usually a set of natural orbitals calculated from an (orbital-unrelaxed)
second-order M{\o}ller-Plesset (MP2) perturbation theory calculation. It is particularly beneficial
to employ an unrestricted Hartree-Fock calculation as a reference, which leads to the incorporation
of symmetry breaking information into the natural orbital construction. Suitable thresholds are
used to select a subset of the MP2 natural orbitals as an initial active space.
This initial active orbital subset is used for the main part: a fast, but low-accuracy
density matrix renormalization group (DMRG) calculation.
After automatic analysis of the correlation information obtained from the DMRG calculation, one or
multiple suggestions for the final, suitable active spaces are made by the software.
The guiding principle behind the design of the ASF is that the majority of its suggestions should
lead to converging CASSCF calculations, with orbitals participating in the correlation within the
active space.

Additional complications are encountered for chemical reactions. Multiple geometries are involved,
typically the reactant and the product, likely the transition state, and possibly an entire
path representing individual steps between the stationary points. The chosen active spaces
need to be reasonable for the respective geometries, but they also need to be consistent, implying
constant size and smoothly evolving orbitals along the reaction path.

The Active Space Finder is built upon PySCF~\cite{Sun2020}, using block2~\cite{Zhai2023} for the
DMRG calculations. While claiming full automatization would be overly ambitious, the ASF makes
active space selection profoundly simpler and more systematic than a manual approach.
It is conceivable to replace the DMRG calculations with other approximate FCI schemes, or one day
even with a low-accuracy calculation on a quantum computer.

While an active space calculation is tailored towards static correlation, it omits the correlation energy involving all electrons
that remain outside the active space. For quantitative purposes it is usually necessary to perform
an additional dynamic correlation calculation building on top of a CASSCF or CASCI reference wave
function. Popular methods include complete active space perturbation theory (CASPT2)~\cite{Andersson1990,Andersson1992},
$n$-electron valence state perturbation theory (NEVPT2)~\cite{Angeli2001,Angeli2002,Angeli2006}
and multi-reference configuration interaction (MRCI)~\cite{Werner1988}.
Moreover, different variants of multi-reference coupled cluster (MRCC) methods have been
published~\cite{Evangelista2018,Lyakh2011}.

Dynamic correlation methods for smaller active spaces often employ internally contracted wave
functions. Only reduced density matrices are needed as the output of the active space solver,
making the methods independent of its specifics. Standard implementations of NEVPT2 or
CASPT2 require four-particle density matrices. The number of tensor elements scales as
$O\left(n_\mathrm{act}^8\right)$ with the number of active orbitals $n_\mathrm{act}$. More
sophisticated dynamic correlation methods may require even higher order density matrices.
This is reasonably unproblematic when the number of active orbitals is small. However, handling
such tensors becomes unfeasible already in conventional calculations with large active spaces.
Thus, the problem of recovering the dynamic multi-reference correlation energy in conjunction with
conventional solvers for large active spaces remain without a universally valid solution.
Suggested approaches are typically strongly tied to the specifics
of the configuration interaction approximation employed, such as methods developed specifically
for DMRG~\cite{Cheng2022,Sokolov2017} or for heat bath configuration interaction~\cite{Blunt2020}.
It seems likely that a method to recover dynamic correlation on top of a quantum computing solver
would depend its specifics, too. We will discuss our own method, that allows a mapping of both active space and remaining
dynamic correlation to quantum computers in the section~\ref{sec:radicals}.

Thus, the preparation steps for active space calculations can employ quantum chemical procedures
that are well-established for conventional calculations.
Needing to choose an active space is a complication that applies to the conventional and quantum
computing cases alike, but it can be mitigated with tools such as the ASF.
The overall assessment in this section is three stars (\starthree).

\subsubsection{Get the job done without the quantum computer}

The most direct assessment of a potential quantum solver is to test its capability in comparison
with conventional configuration interaction solvers for FCI or CASCI. Benchmark references for
large active spaces would be set by existing approximate FCI solver implementations.

One particularly important example is the matrix product state (DMRG) implementation in the
open-source program block2~\cite{Zhai2023}, which is interfaced with the PySCF~\cite{Sun2020}
quantum chemistry package. Capabilities of this combination include FCI, CASCI and CASSCF
calculations. The active space finder (ASF) software makes use of these packages, too.

A different approach to solve the same problem is taken by selected configuration interaction
methods. An example is the iterative-configuration expansion configuration interaction (ICE-CI)
method~\cite{Chilkuri2021_1,Chilkuri2021_2} that is implemented in the proprietary ORCA
package~\cite{Neese2022}. For further examples and references including full CI quantum Monte Carlo
(FCIQMC), adaptive sampling CI (ASCI) or other methods, we refer to literature reviewing the
topic~\cite{Eriksen2020,Baiardi2020}.

Unless the quantum solver guarantees an exact solution in the active space, its accuracy needs to
be benchmarked in addition to its resource requirements. Exact solutions can be calculated conventionally only for relatively small active spaces, and all of the aforementioned methods for larger active spaces introduce approximations. However, the
approximations often make use of adjustable settings, such as the bond dimension in DMRG or the
thresholds in ICE-CI. Changing those parameters permits the user to examine if the
resulting energy is likely to be converged. An additional criterion is provided by the variational
principle, provided it is violated neither by the conventional nor by the quantum solver.

It can be more complicated to assess the quality of the solution for the full system including its
inactive orbitals, especially if dynamic correlation is accounted for. The variational principle does not apply to
methods such as coupled cluster or NEVPT2, nor is it meaningful to compare to DFT total energies.
Additionally, errors in electronic energies from conventional quantum
chemistry methods are usually several orders of magnitude larger than errors in energy differences
between geometries or electronic states. Therefore, a lower total energy does not imply that the 
method is superior in general: it may have worse error cancellation, leading to worse performance
for computed properties. Emphasizing this point is important: total energies can be a useful criterion for approximations to the same method if the variational principle is adhered to, such as different
full CI solvers for identical active orbitals. However, comparing total energies of different methods, such as CASCI with NEVPT2, DFT or coupled cluster, will not lead to meaningful conclusions about their practical performance.

As any computational chemistry method, results calculated on a quantum computer should ultimately
also be benchmarked against experimental results. 
This is the least direct benchmark of the quantum solver itself, as there are many factors at play
besides the electronic structure calculation. At the same time, electronic structure use cases
can prove their usefulness ultimately only if they predict measurable quantities successfully.

Thus, if a quantum computing solver is substituted for a conventional solver in an active space
method, then the same procedures and challenges as for benchmarking conventional methods
would be expected to apply. Of course, this disregards challenges that may be specific to the
quantum solver itself. As mentioned above at HQS we use a DMRG solver implemented in block2~\cite{Zhai2023} and
consider this software implementation to be one of the best availbale complete active space solvers. 
Therefore, we assess our possibility to benchmark against conventional methods with four stars
(\starfour).

\subsubsection{Show quantum advantage}

It is strongly belived that quantum computers will revolutionize quantum chemistry calculations of electronic structure ground and excited states. It is generally argued that quantum computers could achieve exponential speedup over classical methods, particularly for complex chemical systems~\cite{Cao2019QuantumChemQC}. However, a skeptical examination reveals significant hurdles, suggesting that quantum advantage in this domain is not guaranteed. Factors ranging from the nature of molecular systems to algorithmic constraints and competing classical methods might limit the feasibility of transformative quantum gains, especially for ground-state calculations.

One fundamental issue lies in the electronic structure of molecules at equilibrium. Many molecules in their ground states exhibit single-reference character, meaning their wavefunctions are dominated by a single Slater determinant. Classical methods, such as Hartree-Fock, coupled-cluster, or density functional theory (DFT), excel at handling single-reference systems, providing accurate ground-state energies with polynomial computational cost. Quantum computers, by contrast, are expected to shine in multireference scenarios, where strong electron correlation leads to wavefunctions with multiple significant determinants -- a situation common in transition states, radicals (sell also section~\ref{sec:radicals}), or certain transition metal complexes, but less prevalent in equilibrium geometries. 

The standard quantum algorithm for quantum chemistry is quantum phase estimation (QPE). This algorithm and its potential performance has been critically examined in Ref.~\cite{Lee2023Evaluating}. QPE’s success depends on preparing an initial state with polynomial overlap with the true ground state. Lee et al. argue that if such a state can be efficiently prepared, the system’s properties -- such as weak electron correlation or locality -- often enable classical heuristic methods, like variational Monte Carlo, tensor networks, or selected configuration interaction, to solve the problem with comparable efficiency. Conversely, if no initial state with polynomial overlap exists, QPE becomes computationally infeasible due to vanishing success probabilities, precluding quantum advantage. Through numerical studies of model Hamiltonians and ab initio systems, alongside empirical complexity analyses, the authors demonstrate that the conditions facilitating quantum efficiency frequently align with classical tractability. Their conclusion is sobering: exponential quantum advantage is unlikely across broad chemical space, with polynomial speedups being a more realistic outcome. 

An alternative perspective has been described in Ref.~\cite{schleich2025}. This work proposes a quantum framework for simulating entire chemical processes, including reaction dynamics, using a scattering-based approach to state preparation and open-system dynamics with Markovian environments. By simulating time-dependent phenomena like reaction pathways and thermalization, the framework aims for exponential speedup over classical methods. While innovative, this approach is considerably more complex than the static electronic structure calculations central to ground-state quantum chemistry. It involves dynamic processes and environmental interactions, which, while relevant to broader chemical simulations, diverge from the focused task of computing ground-state energies. Clearly the ideas
proposed in Ref.~\cite{schleich2025} are quite interesting, but also fundamentally different from the ideas discussed in this use case section. 

In conclusion, while quantum advantage in ground-state quantum chemistry is theoretically possible, the path forward is fraught with obstacles. The prevalence of single-reference systems, the insights from Ref.~\cite{Lee2023Evaluating} on the interplay between quantum and classical efficiency, and the complexity of dynamic frameworks proposed in Ref.~\cite{schleich2025} highlight the need for tempered expectations. Excited-state calculations offer more promising avenues, but significant research into quantum algorithms, hardware, and benchmarking is required to clarify when and where quantum advantage will materialize. We estimate currently the 
probability to achieve quantum advantage as two stars (\startwo).

\subsection{Radicals}
\label{sec:radicals}

\subsubsection{Identify industry problem}

The singlet-triplet gap, $\Delta E_{ST}$, is one of the important factors defining the chemistry of diradicals, molecules with two unpaired spins~\cite{stuyver2019diradicals}. $\Delta E_{ST}$ is defined as an energy difference between molecule's singlet and triplet states. Typically, diradical are characterized by small values of singlet-triplet gap (a few kcal/mol), which defines their unique properties leading to applications as active materials in photovoltaics~\cite{lukman2017efficient}, photothermal energy conversion~\cite{sun2022diradical} and a luminescent devices~\cite{zhu2025stable}. The ability of a theoretical approach to accurately compute $\Delta E_{ST}$ is instrumental in design of new photo materials and so sets the industrial value of the problem. 

\begin{figure}[!tbp]
  \centering
  \includegraphics[width=0.95\columnwidth]{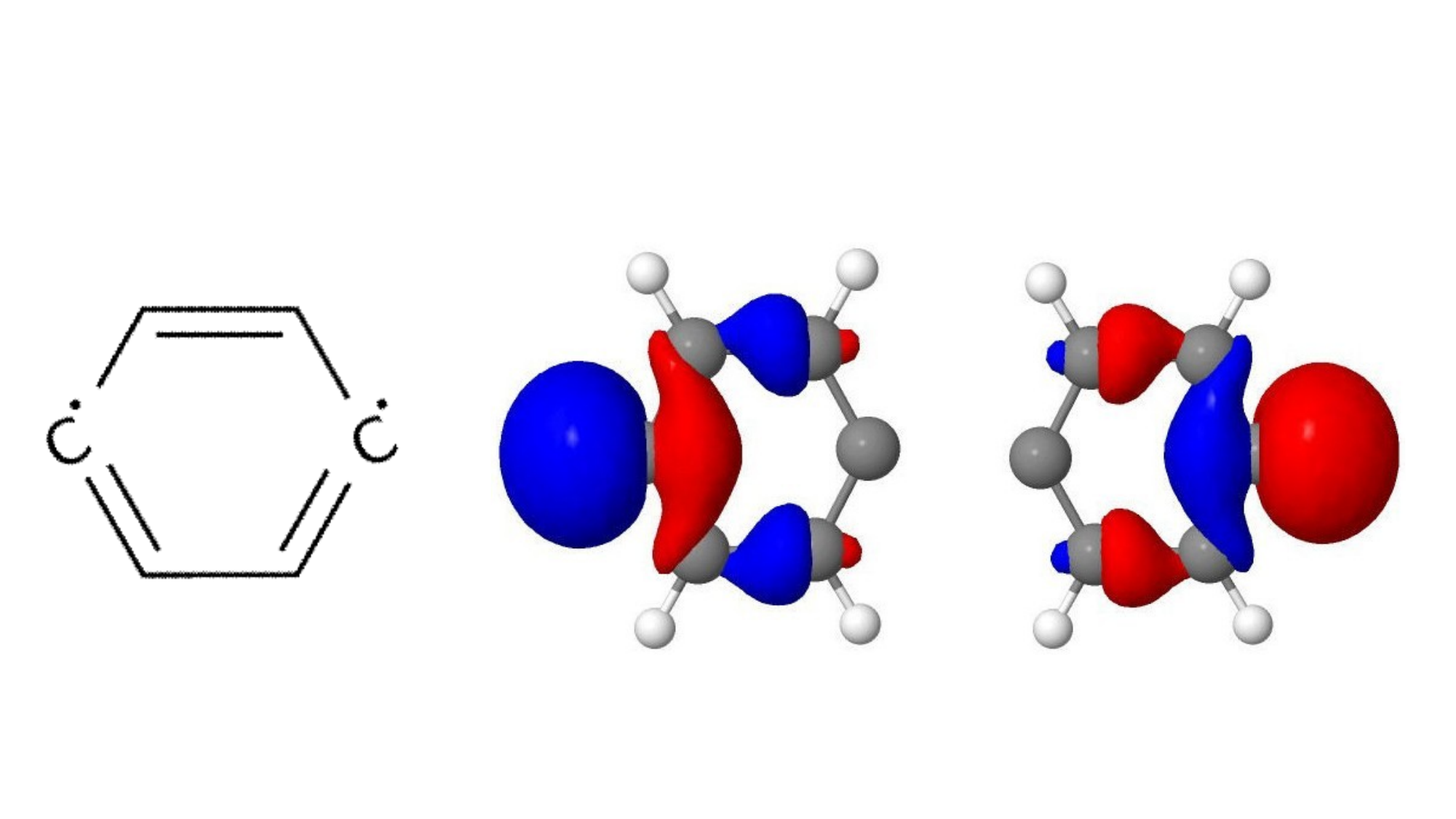}
  \caption{Para-Benzyne, a well known di-radical. A two dimensional description of the structure is shown on the left and the two spin-like orbitals on the right.}
  \label{fig:diradicals}
\end{figure}

In many systems $\Delta E_{ST}$ can be either positive or negative leading to the ground state being singlet or triplet, respectively. Since the triplet-singlet transition is spin forbidden, the spin state of the ground state will define the chemistry and physics of the diradical under normal conditions. A key example is the ability of a system to undergo singlet fission~\cite{smith2010singlet}. This implies a spin-allowed conversion of two molecules in singlet states (ground, $S_0$ and excited, $S_1$) into two low-lying triplet states ($T_1$), which separate from each other:
\begin{equation}
	S_0 + S_1 \rightarrow (T_{1} T_{1})^1 \rightarrow T_1 + T_1.
\end{equation}
Excited singlet states $S_1$ (excitons) are created by light absorption, whereas generation of triplet states $T_1$ is necessary for charge separation. This process underpins many photovoltaic applications in organic materials, its efficiency being dependent on $\Delta E_{ST}$.

There is also a significant academic relevance to studying diradicals because they play a fundamental role in understanding electronic correlation and reactivity in open-shell systems. Diradicals serve as classic examples for testing and developing new electronic structure theories because their electronic properties challenge the assumptions of traditional single-reference quantum chemistry methods. Additionally, diradicals are key intermediates in a variety of important chemical reactions, such as combustion, polymerization, and photochemical processes. Insights gained from research on diradicals enhance our understanding of spin dynamics, electron correlation effects, and nontrivial bonding situations, thereby advancing core concepts in theoretical and computational chemistry.

Here, we propose conventional and quantum solutions of the electronic structure problem for singlet-triplet gap of diradicals using a model of two electrons in two orbitals, augmented with a tailored Random phase approximation (RPA) correction. The relevance of this use case is defined by the two factors: its relevance for chemistry and photomaterials, on the one hand, and the characteristics of the existing quantum chemistry approaches, which are either computationally expensive, or else not robust. To conclude the assessment, we rate relevance of the use case with two stars (\startwo). Such a moderate ranking is underpinned by the fact that the scope is limited to a highly specific property, the singlet-triplet gap, of particular class of compounds, diradicals.  $\Delta E_{ST}$ is essential, but not sufficient for design of new molecules and materials, but more of academic interest. 

\subsubsection{Transform to quantum}

The most characteristic feature of diradicals' electronic structure is the presence of two near-degenerate frontier orbitals. Thus, these systems can be qualitatively described by a two-electrons in two-orbitals model (see Fig.~(\ref{fig:diradicals})), capturing strong (static) electron correlation between the resulting electron configurations~\cite{stuyver2019diradicals}. This implies the use of multireference quantum chemistry methods: they treat static correlation by complete active space self-consistent field methods (CASSCF)~\cite{Olsen2011}, whereas dynamic correlation is often described by perturbation theory. Active spaces consist of all Slater determinants obtained by distributing N electrons into M orbitals. The minimum reasonable active space for diradicals is two electrons in two orbitals ($N=M=2$).  However, a qualitative prediction of such a fine property as the singlet-triplet gap requires taking the dynamic correlation into account in addition to the static one. Standard perturbation theory approaches to treat correlation beyond active space are the complete active space second order perturbation theory (CASPT2)~\cite{Andersson1990} and the N-electron valence state second-order perturbation theory (NEVPT2)~\cite{Angeli2001}. These methods are robust, but in general would require a larger active space than simply the two singly occupied orbitals and exhibit unfavorable exponential scaling of computational cost with system size. The methods to add dynamic correlations would also not map naturally to a quantum computer.    

An alternative computationally cheap approach is to use single-reference approaches. 
However, the ground state of diradical molecules cannot be easily found with these black-box methods unless one exploits unrestricted mean-field approaches breaking the spin symmetry and assuming different orbitals for different spins~\cite{neese2004definition}. These methods are hard to converge to the correct state and they are prone to heavy spin contamination, 
sacrificing correct wave function properties for the sake of decent energetics~\cite{yang2016nature}. 

Random phase approximation (RPA)~\cite{bohm1951collective, pines1952collective, bohm1953collective}, a work horse of electronic structure theory for materials and molecules~\cite{eshuis2012electron}, has been applied to compute singlet-triplet gaps with considerable success. Different flavours of single-reference (\textit{i.e.} applied to a single Slater determinant) RPA captures dynamic correlation and is able to accurately compute singlet-triplet gaps for weakly correlated systems~\cite{Dhingra_RPA}. For these, $\Delta E_{ST}$ are significantly larger than for the diradicals making those materials less relevant for the aforementioned applications in energy and photonics. At the same time RPA-based approaches can provide accurate singlet-triplet gaps for a particular class of organic chromophores -- linear oligoacenes~\cite{yang2016nature, Dhingra_RPA}, exhibiting significant diradical character. However, there is little evidence that single-reference RPA methods can accurately describe electronic structure of arbitrary diradicals.

To solve for singlet-triplet gaps of diradicals one needs a second-quantized quantum chemical Hamiltonian \cite{PurpleBook}:
\begin{equation} \label{eq:Hamiltonian}
H = \sum_\sigma\sum_{pq} t_{pq} c^\dagger_{p\sigma} c_{q\sigma} 
+ \frac{1}{2} \sum_{\sigma\sigma'} \sum_{pqrs} h_{psqr} c^\dagger_{p\sigma} c^\dagger_{q\sigma'} c_{r\sigma'} c_{s\sigma}\end{equation}\\
where indices ${p,q,r,s}$ run over spatial orbitals; $t_{pq}$ and $h_{pqrs}$ are one-electron and two electron reduced density matrix (1-RDM and 2-RDM) elements in the molecular orbital basis, respectively; $t_{pq}$ and  $h_{pqrs}$ are one-electron integrals and two-electron integrals, respectively. The operators $c^\dagger_{p\sigma}$ ($c_{p\sigma}$) create (annihilate) an 
electron in orbital $p$ with spin $\sigma$.

In our approach, the same Hamiltonian is used for both singlet and triplet states, absolute total energies of which are not computed, but only their difference. This implies that both states have very similar equilibrium structures, which is a major approximation, that often holds. Whereas, one- and two-electron integrals are explicitly used in the calculation, the 1-RDM density matrix is used to define the two active orbitals. In technical terms, to prepare the data for our classical and quantum algorithms for $\Delta E_{ST}$ one needs:
\begin{enumerate}
	\item Optimized equilibrium structure for a singlet or a triplet state. This is obrained by using geometry optimization with a proper CASSCF wave function, using a minimum (2,2) or extended active space.
	\item Hamiltonian from CASSCF with the minimum or extended active space. The Hamiltonian can be computed for each of the states, or for a state-averaged wave function. This must ideally include all orbitals, rather than being restricted to an active space.
\end{enumerate}

CASSCF calculations including geometry optimization are computationally demanding. However, with a minimum or small active space such calculations are accessible, especially in comparison with multireference wave functions, such as NEVPT2 or CASPT2, needed for accurate systematic calculations of singlet-triplet gaps. Another technical challenge is the input/output operations of electron integral tensors, $t_{pq}$ and  $h_{pqrs}$. Those are computed by standard quantum chemistry packages and can be stored on hard drive. For small- and medium-sized systems writing and reading in of those quantities is possible in terms of memory and speed if permutation symmetry of the integrals is utilized. We use PySCF program~\cite{Sun2020} for CASSCF calculations, where the latter can be accelerated by density matrix renormalization group theory (DMRG) implementation of the CAS solver in the block2 program~\cite{Zhai2023}. PySCF uses the libint library \cite{Libint2} for analytical evaluation of electron integrals over standard atomic Gaussian basis functions, which are stored as native numpy arrays \cite{harris2020array} via the Python interface of PySCF.

After this procedure in which we used standard methods from quantum chemistry, we now perform transformations on the level of the Hamiltonian, to use the classically efficent static approximation to solve the random phase approximation, but also allow for a higher order approximation when mapping to a quantum computer. Since here we are considering diradicals, 
we separate the two orbitals hosting the two unpaired electrons and number them as the orbitals $1$ and $2$. We will call this the active space from now on. 
All remaining orbitals, both doubly occupied and empty, are numbered as $3,4,5,\dots, N$. 
These orbitals form the environment for the active space.
After such separation, the Hamiltonian  takes the form
\begin{equation}
H = H_{12} + H_{\rm env} + V,
\end{equation}
where $H_{12}$ is the Hamiltonian of the active space,
$H_{\rm env}$ is the Hamiltonian of the environment. This basically means both Hamiltonians, take the same form as 
equation (\ref{eq:Hamiltonian}), but with restricted summation. The Term $V$ is the coupling between the two subsectors of the Hamiltonian and contains all terms that have mixed indices. 

In the RPA approximation every excitation from a doubly occupied orbital $\alpha$, lying below the Fermi level, to an empty orbital $m$ above the Fermi level,
i.e. the excitation $\alpha\to m$, is replaced by an oscillator. On the operator level, the following replacement is made:
\begin{equation}
c_{m\uparrow}^\dagger c_{\alpha\uparrow} + c_{m\downarrow}^\dagger c_{\alpha\downarrow} \rightarrow  \sqrt{2} a_{m\alpha}^\dagger ,
\end{equation}
\begin{equation}
c_{\alpha\uparrow}^\dagger c_{m\uparrow} + c_{\alpha\downarrow}^\dagger c_{m\downarrow} \rightarrow  \sqrt{2} a_{m\alpha},
\end{equation}
where $a_{m\alpha},a_{m\alpha}^\dagger$ are the ladder operators of the oscillator. 
The frequency of such oscillator corresponds to the difference between the Hartree-Fock energies of the excited and the ground states
\begin{equation}
\hbar\omega_{m\alpha} =  \tilde t_{mm} - \tilde t_{\alpha\alpha} - h_{mm\alpha \alpha } + h_{m\alpha m\alpha},
\end{equation}
where
\begin{equation}
\tilde t_{pq} = t_{pq} + \sum_{k=3}^N (2h_{pqkk}- h_{pkkq})n_k
\end{equation}
are the hopping amplitudes between the orbitals of the environment modified by Hartree-Fock interaction within the environment itself,
and the occupation numbers of the orbitals are $n_q=1$ for the doubly occupied orbitals and $n_q=0$ for the empty ones. 
One can verify that the RPA approximation is valid if
\begin{subequations}
\begin{align}
h_{m\alpha m\alpha} &\ll \hbar\omega_{m\alpha}, \\
h_{mmmm} + h_{\alpha\alpha\alpha\alpha} - 2h_{mm\alpha\alpha} &\ll \hbar\omega_{m\alpha}
\end{align}
\end{subequations}
for all possible pairs of the doubly occupied and empty orbitals of the environment $m,\alpha$. The first of these conditions requires weak exchange interaction
between the environment orbitals, as expected. 
Thus, in the direct RPA approximation the Hamiltonian reads
\begin{equation}\label{eq:Final_RPA_Hamiltonian}
 H \approx H_{\rm RPA} = H_{12} + H_{\rm env} + V_{\rm RPA}.
\end{equation}
with the transformed Hamiltonian of the environment
\begin{align}
H_{\rm env} =\; & E_{\rm env}^{\rm HF} \notag \\
& + \sum_{m,\alpha} \left( \hbar\omega_{m\alpha}\, a^\dagger_{m\alpha}a_{m\alpha} - h_{m\alpha m\alpha} \right) \notag \\
& + \sum_{m\alpha,n\beta} h_{m\alpha n\beta} (a^\dagger_{m\alpha} + a_{m\alpha}) (a^\dagger_{n\beta} + a_{n\beta}),
\end{align}
and the transformed interaction term $V_{\rm RPA}$,
\begin{equation}
V_{\rm RPA} = \sqrt{2} \sum_{\sigma=\uparrow,\downarrow}\sum_{m\alpha}\sum_{p,q=1}^2 h_{pqm\alpha} (a^\dagger_{m\alpha} + a_{m\alpha}) c^\dagger_{p\sigma} c_{q\sigma}.
\end{equation}
The Hartree-Fock energy of the environment orbitals, appearing in $H_{\rm env}$, is given by the sum
\begin{equation}
E_{\rm env}^{\rm HF} = 2\sum_{q=3}^N t_{qq} n_q + \sum_{pq=3}^N ( 2h_{ppqq} - h_{pqqp} )n_pn_q.
\end{equation}
The Hamiltonian (\ref{eq:Final_RPA_Hamiltonian}) can be seen as the final product of our \textit{Transform to Quantum} efforts.
From this point onward we can now try to solve the problem classically, which we will discuss in the next section, or we can try
to map the problem to the quantum computer which will be discussed in section~\ref{subsubsec:Radicals_Show_Quantum_Advantage}. The derivation of the Hamiltonian has also been discussed in detail in Ref.~\cite{Shirazi_RPA}.

The overall assessment of ability to transform to quantum is three stars (\starthree). The main difficulties are the assumption of the same geometric structure for both singlet and triplet states, as well as the relatively complex quantum chemistry calculations required to generate the input Hamiltonian.

\subsubsection{Get the job done without the quantum computer}

Quantum computing approaches to compute singlet-triplet gaps in diradicals can be benchmarked against either conventional general-purpose electronic structure methods or a custom RPA approximation designed at HQS Quantum Simulations for computing this particular property \cite{Shirazi_RPA}. 

Conventional methods include highly accurate and expensive methods based on the CASSCF reference (multireference methods): perturbation theory variants CASPT2 and NEVPT2, described above, as well as other multireference variants of coupled-cluster (MRCC) \cite{lyakh2012multireference} and configuration interaction (MRCI) \cite{szalay2012multiconfiguration} theories. A cheaper and way less robust alternative are spin-symmetry-broken (unrestricted) single-configurational methods (density functional theory \cite{neese2004definition}).

An RPA model for singlet-triplet gaps in diradicals developed at HQS \cite{Shirazi_RPA} is based on the two-electrons-in-two orbitals model, which captures the most essential strong correlation in a diradical. The effect of empty and doubly occupied orbitals (dynamic correlation) is taken into account by RPA. In the static limit, direct RPA approximation leads to the renormalization of the parameters of the two-orbital model. Electron interactions (one- and two-electron integrals) are taken from the minimum CASSCF-(2,2) conventional quantum chemistry calculation. The computational price of such simulation is marginally larger than that of a single-configurational one, whereas the RPA correction has an exact analytic solution, adding practically no computational overhead.

\begin{figure}[!tb]
	\centering
	\includegraphics[width=1.0\columnwidth]{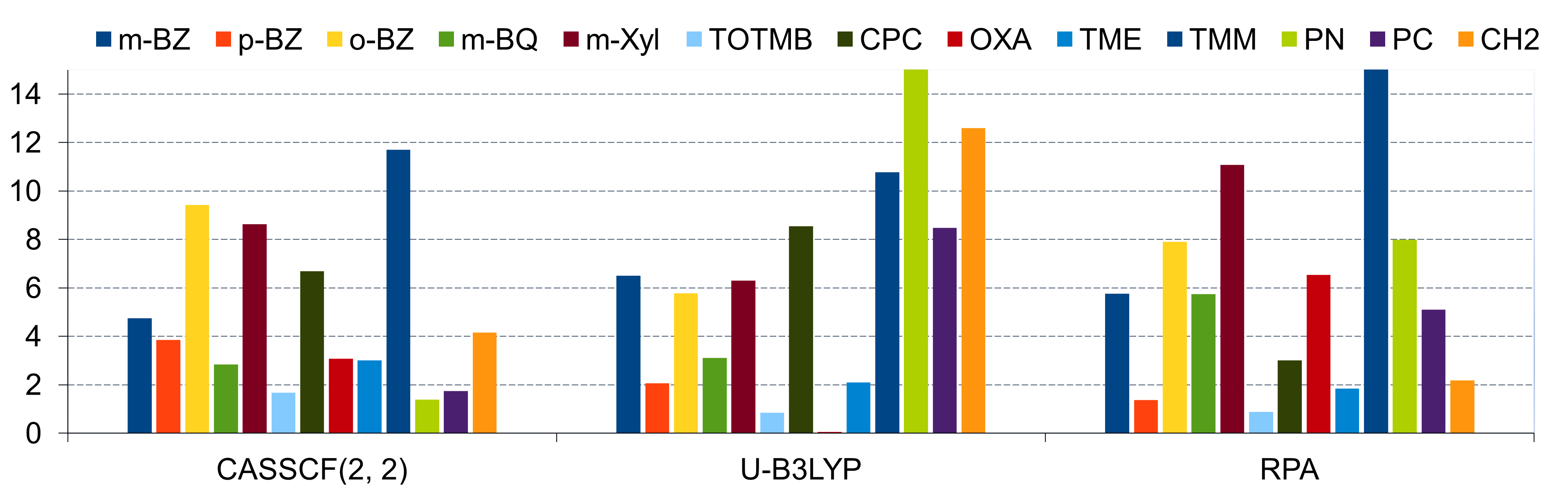}
	\caption{Errors in $\Delta E_{ST}$, in kcal/mol, for various computational methods as compared to the multireference coupled-cluster results. Dataset consists of 13 organic molecules of different diradical character.}
	\label{fig:errors}
\end{figure}

Figure~\ref{fig:errors} shows accuracy of our RPA approach as compared to the multireference methods as well as density functional theory (with B3LYP functional) in the unrestricted (spin-symmetry-broken) variant. RPA performs on average better than density function theory methods and is commensurate with CASSCF-(2,2) based on the same selected active space. However, CASSCF, and also NEVPT2, calculations with larger active spaces exhibit consistently better accuracy, at a price of larger computational costs.

We rate this use case as two stars (\startwo) since there is a wide selection of general conventional electronic structure methods varying in computational cost, accuracy and robustness, as well as an efficient custom HQS developed RPA approach, demonstrating reasonable results at moderate costs. There are still many steps that can be improved for classical calculations to achieve a better rating over time.

\subsubsection{Show quantum advantage}\label{subsubsec:Radicals_Show_Quantum_Advantage}

With Hamiltonian (\ref{eq:Final_RPA_Hamiltonian}) we have constructed a minimal active space consisting of 2 electrons in 2 orbitals, resulting in four possible electronic states. This allows the electronic structure of the active space to be fully described by a $4\times 4$ Hamiltonian matrix. The remaining orbitals, which are not part of the active space, have been transformed into bosonic modes using the random phase approximation, effectively modeling them as a bosonic bath. Consequently, the system can be described as a 4x4 electronic system coupled to a bosonic bath, analogous to the spin-boson model but with a four-level system instead of a two-level (spin) system.

The spin-boson model, where a two-level system is coupled to a bosonic bath, has been extensively studied in the context of quantum computing and quantum simulation. Numerous approaches have been developed to solve its dynamics, including digital quantum simulation on platforms like IBM quantum computers, variational quantum algorithms, and analog simulations using trapped ions or superconducting circuits. These methods address challenges such as modeling dissipative dynamics, handling strong system-bath coupling, and simulating structured spectral densities~\cite{Burger2022Digital, Sun2025SpinBosonNatCommun}.

Given the structural similarity between the spin-boson model and the 4x4 system coupled to a bosonic bath we have constructed, these established techniques should be adaptable to our situation. The primary difference lies in the dimensionality of the system (4x4 versus 2x2), which increases the complexity of the Hamiltonian but does not fundamentally alter the nature of the system-bath interaction. Quantum algorithms, such as Trotterization or variational methods, can be extended to handle the larger Hilbert space of the 4x4 system, while analog platforms could be configured to represent four-level systems coupled to bosonic modes. 

Of course there are also differences to the existing work on the spin boson model. In the existing work, often well known bosonic
spectral functions are being simulated. The spectral function for a molecule obviously looks quite different from an ohmic bath. So there are still many points to analyze. However, the key point of many of these efforts is that it is possible to map spin-boson model to a quantum processor with quite low depth, and this should also be the case in our situation. While this offers interesting opportunities, there is still a lot of work to be done to solve for the singlet triplet splitting within the 
random phase approximation on a quantum computer. Nonetheless, as compare to the direct approach using active space methods, we directly integrate dynamic correlations in this approach, generate an extremly effective mapping to the quantum computer and are looking for energy differences in the spectrum of a Hamiltonian, instead of ground state properties. Therefore we rank the current probability to achieve quantum advantage for this use case at \starthree.

\section{Conclusions}
\label{sec:conclusions}

\begin{table*}[!htbp]
  \centering
  \renewcommand{\arraystretch}{1.2}
  \rowcolors{2}{gray!30}{white}
  \begin{tabular}{
      @{\hspace{0.15cm}}p{6.4cm}@{\hspace{0.15cm}}|
      @{\hspace{0.15cm}}p{3.2cm}@{\hspace{0.15cm}}|
      @{\hspace{0.15cm}}p{3.2cm}@{\hspace{0.15cm}}|
      @{\hspace{0.15cm}}p{3.2cm}@{\hspace{0.15cm}}
  }
  \rowcolor{gray!30}
  \textbf{Criteria}
      & \centering NMR
      & \centering Multireference Chemistry
      & \centering Radicals
      \tabularnewline \hline
  \raggedright\textcolor{hqstext}{\textbf{Identify Industry Problem}}
      & \centering\starfour
      & \centering\starthree
      & \centering\startwo
      \tabularnewline
  \raggedright\textcolor{hqstext}{\textbf{Transform to Quantum}}
      & \centering\starfour
      & \centering\starthree
      & \centering\starthree
      \tabularnewline
  \raggedright\textcolor{hqstext}{\textbf{Get the Job Done Without the Quantum Computer}}
      & \centering\starfive
      & \centering\starfour
      & \centering \startwo
      \tabularnewline
  \raggedright\textcolor{hqstext}{\textbf{Show Quantum Advantage}}
      & \centering\starfour
      & \centering\startwo
      & \centering \starthree
      \tabularnewline
  \end{tabular}
  \caption{HQS use-case classification and ratings. The table categorizes use cases based on four key criteria: identification of a real industry problem, pre- and post-calculations to transform to quantum, benchmarking against conventional methods without the quantum computer, and quantum algorithm selection that has the potential for quantum advantage. Each criterion is rated using a star system: \starone\ (Initial understanding or academic relevance), \starthree\ (Some real-world validation or industry-related insights), and \starfive\ (Fully validated with industry proof and rigorous benchmarking). A five-star rating in the first three categories requires a five-star rating in benchmarking (``Get the Job Done Without the Quantum Computer"').}
  \label{tab:qa-use-cases}
  \end{table*}

By assessing quantum computing use cases through the four interdependent criteria -- Identify industry problem, Transform to quantum, Get the job done without the quantum computer, and Show quantum advantage -- we provide a structured perspective on evaluating their relevance and potential. In conclusion, we briefly recap our assessment of the three use cases presented in this work, as summarized in Table~\ref{tab:qa-use-cases}. These ITBQ criteria  -- \textbf{I}dentify, \textbf{T}ransform, \textbf{B}enchmark, Show \textbf{Q}uantum Advantage -- laid out do not just serve as a static assessment instrument; they function as a dynamic checklist for ongoing development. They help highlight areas where further work is needed, guide product and market research, and support the setting of priorities in both technology development and industry engagement. In this way, the scheme enables transparent communication, targeted innovation, and continuous improvement in the pursuit of quantum advantage.

Nuclear Magnetic Resonance is a firmly established analytical technique, and the calculation of realistic spectra is feasible, provided that suitable NMR parameters are available or can be computed. Nevertheless, the need for additional effort to ensure seamless integration of simulations in the established analytical workflow justifies not awarding the maximum rating. Pre- and post-processing procedures to estimate NMR parameters using quantum chemistry or data-driven approaches are well-established, but computed NMR parameters cannot fully agree with experiment, and certain factors, such as the impact of solvents or interactions between solute molecules, are very difficult to model. Integration of NMR simulations into analysis solutions or new software remains use case dependent. For benchmarking, advanced classical solvers are continuously developed by HQS. The quantum advantage is seen as highly promising, as this use case involves time evolution and solving the Schrödinger equation, which is naturally suited for quantum computers. But classical solvers are clearly highly competitve and clustering can in many cases provide highly accurate approximations.

The assessment of Multireference Chemistry focuses on use cases characterized by statically correlated systems that require a large active space for qualitatively correct treatment. This is relevant for applications such as excited electronic states and transition metal systems, as discussed in the literature, demonstrating a clear academic interest. The pre-processing steps are based on quantum chemical procedures that are established for conventional calculations; however, active space selection is a complication shared by both conventional and quantum computing approaches and can be mitigated by the Active Space Finder. A central unresolved issue is how dynamic correlation would be recovered with a quantum solver in large active spaces. When a quantum computing solver replaces the conventional solver, identical procedures and benchmarking criteria apply, though quantum-specific challenges may arise. The question of quantum advantage remains open and requires further investigation.

In the case of Radicals, the assessment considers both the relevance for chemistry and photomaterials and the characteristics of the current quantum chemistry approaches, which tend to be either computationally expensive or insufficiently robust. The given rating reflects the limited scope of the use case, as it targets a highly specific property, the singlet-triplet gap, of a particular class of compounds, namely diradicals. While this gap is an important parameter, it alone is not sufficient for the design of new molecules and materials. Pre- and post-processing has the main challenges being the assumption of the same geometric structure for singlet and triplet states, and the complex quantum chemistry computations needed to generate the input Hamiltonian. That said, it is broadly agreed that even achieving quantum advantage for purely academic research is a worthwhile goal.  For benchmarking we note the availability of a broad range of conventional electronic structure methods that differ in computational cost, accuracy and robustness, alongside an efficient custom HQS-developed approach producing reasonable results at moderate costs. Quantum advantage has not yet been assessed in detail and remains to be explored further. But the specific transformations we have chosen allows for interesting possibilites when it comes to mapping the problem to the quantum computer.   

The evaluation of quantum use cases is essential for directing resources efficiently and ensuring that both time and investments are allocated to the most promising opportunities. Well-defined assessment criteria help stakeholders identify real value and potential market impact. This aspect is particularly important at the current stage of quantum computing, where many projects are publicly funded and there is a clear responsibility not to waste resources unnecessarily. Structured and transparent evaluation supports informed decision-making and accountability for the chosen development paths.

Without a systematic assessment, there is little basis for investment and strategic planning. The absence of clear criteria increases the risk of repeating past mistakes and missing important opportunities or failing to recognize lessons learned. In practice, every organization implicitly evaluates use cases, but explicit, standardized criteria enable more robust discussions and knowledge transfer.

Once established, such an evaluation scheme serves as a practical tool for learning, prioritization, and project management. It enables teams to focus their efforts where quantum advantage can realistically be achieved, guiding the planning and allocation of resources across different projects and use case domains. This approach helps to systematically drive progress towards quantum advantage, as illustrated by advanced cases such as NMR.

\section*{Acknowledgments}

We thank Sonia Álvarez Barcia, Nicklas Enenkel, Peter Schmitteckert, Reza Shirazi, Florian Wullschläger for their insightful discussions and valuable contributions to the creation of the graphics used in this publication. We gratefully acknowledge PhytoLab GmbH \& Co. KG for sharing the experimental data employed to illustrate NMR applications. This work was supported by by the Federal Ministry of Research, Technology and Space (BMFTR) through the QSolid project (Grant no. 13N16155) and  through the PhoQuant (Grant no. 13N16107) project. 

\bibliography{references,nmr,asf,radicals}

\end{document}